\pdfoutput=1 %

\documentclass[sigconf,nonacm]{acmart}
\usepackage{graphicx}
\usepackage[table]{xcolor}
\usepackage{soul}
\usepackage{xspace}
\usepackage{tabularx}
\usepackage{makecell}
\usepackage{subcaption}
\usepackage{booktabs}
\usepackage{multirow}
\usepackage{multicol}
\usepackage{enumitem}
\usepackage[nameinlink,noabbrev]{cleveref}
\usepackage{listings}
\newcommand{\name}{\textsc{VeilLens}\xspace}

\newcommand{\todo}[1]{} %

\begin{document}

\title{\name: Explainable and Granular Analysis and Protection Against Visual Privacy Risks Beyond Visible Appearance}

\author{Weihao Jin}
\authornote{Both authors contributed equally to this research.}
\authornote{Corresponding author.}
\orcid{0009-0005-5070-2283}
\affiliation{%
  \institution{University of Michigan-Ann Arbor}
  \city{Ann Arbor}
  \state{Michigan}
  \country{USA}
}
\email{jackyjin@umich.edu}

\author{Zuchen Li}
\authornotemark[1]
\orcid{0009-0007-4244-4354}
\affiliation{%
  \institution{University of Michigan-Ann Arbor}
  \city{Ann Arbor}
  \state{Michigan}
  \country{USA}
}
\email{zuchenli@umich.edu}

\author{Ke Sun}
\orcid{0000-0001-9563-1398}
\affiliation{%
  \institution{University of Michigan-Ann Arbor}
  \city{Ann Arbor}
  \state{Michigan}
  \country{USA}
}
\email{kesuniot@umich.edu}


\begin{abstract}
Recent vision-language models (VLMs) have substantially elevated the threat level of visual privacy. A single shared image can reveal sensitive information not only through explicit identifiers, but also through weak and compositional cues that enable semantic reasoning about identity, precise locations, and social relationships.
Existing visual privacy tools primarily target predefined visible identifiers or apply generic image transformations, making them ill-suited for reasoning-capable VLM attackers and offering limited explainability, granularity, and user control.
We present \name{}, an explainable framework that identifies and grounds privacy-relevant cues, protects them through controllable generative editing, and verifies residual leakage in an iterative reasoning--grounding--editing--verification loop.
We evaluate \name{} on $1{,}454$ images with more than $20{,}000$ VLM API calls across compositional privacy, geolocation, and social-relationship inference. \name{} reduces mean inferred privacy risk by $48.7\%$ on $500$ compositional-privacy images and block-level ($\leq 0.5$ km) geolocation success from $40.2\%$ to $14.6\%$ under Gemini-3.1-Pro-Preview. 
A 25-participant user study shows improved privacy awareness while preserving image utility, with $70.7\%$ of responses preferring the protected images.
\end{abstract}
\begin{teaserfigure}
  \centering
  \includegraphics[
    width=1\textwidth, clip
  ]{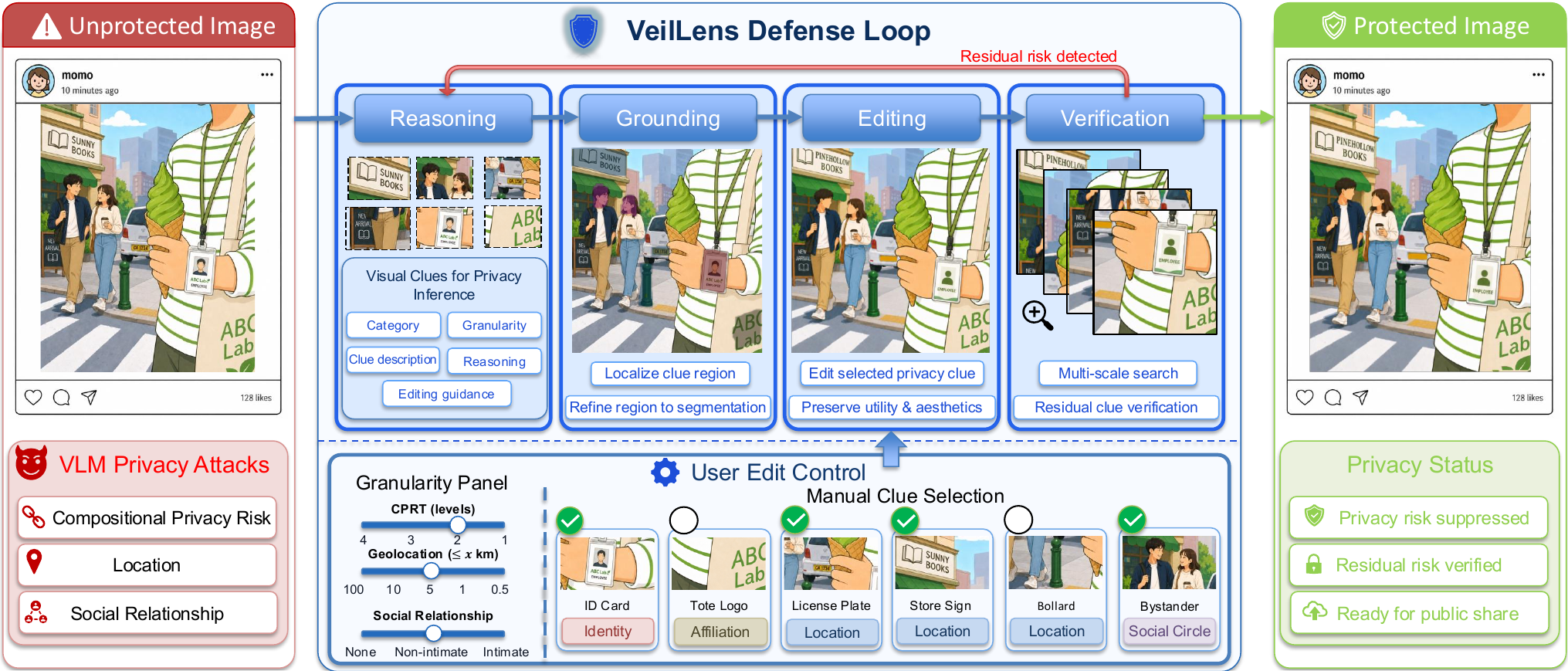}
  \caption{
  Given an unprotected image and user-specified protection preferences, \name identifies visual clues supporting privacy attacks.
  It then localizes clues, edits user-selected evidence, and verifies residual leakage through an iterative defense loop, producing a protected image that suppresses sensitive inference while remaining suitable for public sharing.
  }
  \Description{System overview showing an original social-media-style image on the left and a protected version on the right. Between them, the VeilLens defense loop contains four stages: reasoning, grounding, editing, and verification. The reasoning stage identifies privacy clues and associates them with privacy categories and granularities. Grounding localizes the clues in the image, editing modifies user-selected regions while preserving benign content, and verification searches for residual clues and can trigger another iteration. A control panel illustrates protection granularities for compositional privacy, geolocation, and social relationships, together with manual clue selection. The output panel indicates that privacy risk has been suppressed and residual risk verified.
  }
  \label{fig:teaser}
\end{teaserfigure}

\maketitle

{\let\thefootnote\relax\footnotetext{Preprint.}}

\section{Introduction}

Photo sharing has become a routine global practice: billions of images are
uploaded each day to platforms such as Instagram, X, and
Facebook~\cite{ahern2007overexposed}. Yet these shared images can expose
individuals to privacy risks that extend far beyond explicit identifiers such
as faces and license plates~\cite{acquisti2015privacy}.
A single photograph can reveal geographic location, daily routines, social
relationships, and institutional affiliations through implicit visual cues that users
may neither intend to disclose nor recognize as
sensitive~\cite{zhang2022privacy, ahern2007overexposed, kokolakis2017privacy}.

Unfortunately, the rapid deployment of frontier VLMs into everyday applications
has further altered this threat landscape~\cite{openai2024gpt4o,
openai2025gpt52, geminiteam2025gemini25, google2025gemini3flash,
google2025gemini3pro}.
The advanced reasoning capabilities that make these models useful also enable
powerful privacy inference: a low-effort adversary can upload a victim's photo
and obtain detailed inferences that previously required expertise or specialized
pipelines.
Recent work shows that VLMs can infer precise locations from weak visual signals
through compositional reasoning~\cite{liu2025mission}, and recover identity or
sensitive attributes from contextual cues that users may neither recognize nor
aggregate~\cite{liu_eye_2025, tsaprazlis_rethinking_2026}.
\emph{As a result, visual privacy protection must move beyond detecting and
obscuring predefined visible identifiers to also account for compositional cues,
helping users understand, control, and mitigate the broader privacy risks exposed
by ordinary shared images.}

Existing visual privacy defenses fall short along three axes.
First, \textbf{coverage and grounding}: many target predefined risks such as faces, PII, or geolocation, overlooking leakage from combinations of contextual cues and offering limited evidence for the resulting inference~\cite{tsaprazlis_rethinking_2026}.
Second, \textbf{user steering}: existing defenses provide limited support for jointly controlling protection granularity and individual edits, making it difficult for users to navigate the privacy--utility trade-off~\cite{shu_cardea_2018,monteiro_imago_2025}.
Third, \textbf{verifiability}: most defenses are one-shot or evaluated through human studies and task-specific CV models, while iterative assessment of residual leakage under VLM threats remains limited~\cite{abdulaziz_evaluation_2025,aufschlager_towards_2026}.

To address these limitations, we propose \name{}, an explainable and
user-controllable visual privacy protection framework against reasoning-capable VLM adversaries. 
As shown in Fig.~\ref{fig:teaser}, 
rather than treating privacy protection as
a one-shot sanitization problem, \name{} formulates it as a
\emph{reasoning--grounding--editing--verification} loop:
given an input image and a user-specified privacy objective, \name{} uses a
trusted VLM oracle to reason about what sensitive information may be inferred
and why, grounds the supporting visual evidence as editable regions, applies
user-controllable generative inpainting to selected clues, and iteratively
verifies whether the protected image still leaks the target attribute.

Building \name{} introduces three key challenges.

\emph{First, how can \name{} accurately reason about and ground potential
privacy leakage while remaining explainable to users?}
Although a VLM oracle can identify privacy risks at a high level, it often
does not localize the specific visual evidence responsible, limiting
transparency and user trust.
To address this, \name{} decouples reasoning from localization: a trusted VLM oracle identifies
privacy-relevant cues in language, which are then grounded into pixel-level
segmentation masks through an open-set detection-and-segmentation pipeline,
exposing the evidence chain to the user.

\emph{Second, how can \name{} help users navigate the privacy--utility 
trade-off when sensitive inferences arise from many weak contextual cues?} 
Different users and sharing contexts entail different privacy protection 
preferences, making a one-size-fits-all removal strategy inappropriate.
To make this trade-off explicit, \name{} allows users to specify a target 
protection granularity and presents each inferred cue with its associated 
risk and localized mask for selection. 
Together, these representations allow users to weigh the 
privacy risk of each clue against the visual content affected by its removal, 
enabling informed edit decisions.
\emph{Third, how can \name{} verify protection against strong VLM-based
adversaries?}
A single pass of reasoning and editing may not eliminate all leakage: a VLM
oracle tends to surface only the most salient cues, while a VLM adversary can
still exploit residual contextual signals to recover the target attribute.
To address this, \name{} introduces an iterative verification loop: after each
editing round, an oracle VLM re-evaluates the protected image and reports
residual evidence. \name{} grounds the reported cues and performs additional
inpainting until the target inference is suppressed.

We implement \name{} as an end-to-end pipeline 
integrating a frontier VLM oracle~\cite{google2025gemini3pro, openai2025gpt5}, an open-set detector and
segmentation model, and a diffusion-based inpainting backend.
We evaluate \name{} at scale across $1{,}454$ images, involving more than
$20{,}000$ VLM API calls, four frontier VLM adversaries, five recent
defense baselines and three privacy threat settings.
For general compositional privacy, \name{} reduces the mean CPRT privacy
risk score~\cite{tsaprazlis_rethinking_2026} from $0.314$ to $0.161$, a $48.7\%$ relative reduction.
For geolocation, \name{} reduces block-level ($\le 0.5\,km$) localization success from
$40.2\%$ to $14.6\%$ under Gemini-3.1-Pro-Preview, and achieves a $1.72\times$ lower attack
success rate than the strongest prior geolocation-specific defense~\cite{liu_geoshield_2026}.

It also outperforms direct generative editing with GPT-Image-2 ($17.4\%$) at this
threshold, while additionally providing clue-level attribution, granularity control, and
iterative verification.
For social-relationship privacy, \name{} reduces VLM relationship inference accuracy by over $92\%$ across four frontier attackers \cite{li_dual-glance_2017}.
A user study with $N=25$ participants further shows the privacy and utility trade-off of \name{}.
$70.7\%$ ($106/150$) of responses prefer the protected image, and $52\%$
($13/25$) report willingness to use \name{} in everyday sharing today.

Our key contributions are as follows:

$\bullet$ We introduce a visual privacy protection framework against visual privacy risks beyond visible appearance. It provides granularity-aware, clue-level user steering protection while preserving image utility.

$\bullet$ We introduce an iterative verification loop using a VLM oracle to achieve better privacy guarantee, which verifies protected images after editing and drives further targeted inpainting when residual leakage remains.

$\bullet$ We conduct a comprehensive evaluation of \name{} across three privacy threat settings, demonstrating
that \name{} improves privacy protection and user awareness while preserving a
favorable privacy--utility trade-off.

\begin{table*}[t]
\centering
\caption{Comparison of prior visual privacy defenses and \name{} across protection scope, defense target, and key capabilities.}
\label{tab:comparison}
\small
\resizebox{\textwidth}{!}{%
\begin{tabular}{lccccccc}
\toprule
Method 
& \shortstack{Visual\\privacy threat} 
& \shortstack{Defense\\target} 
& \shortstack{Reasoning-aware\\defense} 
& \shortstack{Granularity-aware\\User steering} 
& \shortstack{Clue\\explainability} 
& \shortstack{Utility\\preserving} 
& \shortstack{Verification\\loop} \\
\midrule

MM-PGD \cite{le_rethinking_2022} 
& Geolocation 
& CV models 
&  
&  
&  
&  
&  \\

ADGAN \cite{xiong_adgan_2021} 
& Geolocation 
& CV models 
&  
&  
&  
& \checkmark 
&  \\

LocGuard \cite{ma_locguard_2024} 
& Geolocation 
& CV models 
&  
&  
&  
&  
&  \\

ReasonBreak \cite{zhang2026disruptinghierarchicalreasoningadversarial} 
& Geolocation
& VLM
& \checkmark
&
&
& \checkmark
&
\\

GeoShield \cite{liu_geoshield_2026} 
& Geolocation 
& VLM 
& \checkmark 
&  
& \checkmark 
&  
&  \\

\midrule

SVIA \cite{liu_svia_2025} 
& PII 
& CV models 
&  
&  
&  
& \checkmark 
&  \\

Redaction \cite{orekondy_connecting_2018} 
& PII 
& Human, OCR 
&  
& 
&  
& \checkmark 
&  \\

Cardea \cite{shu_cardea_2018} 
& PII 
& Human, CV 
&  
& \checkmark 
& \checkmark 
&  
&  \\

Aragorn \cite{venugopalan_aragorn_2023} 
& PII 
& Human, CV 
&  
& \checkmark 
& \checkmark 
&  
&  \\

Unsafe2Safe \cite{dinh2026unsafe2safecontrollableimageanonymization} 
& PII
& General
& \checkmark
&
& \checkmark
& \checkmark
&
\\
 
CAIAMAR \cite{aufschlager_towards_2026} 
& PII 
& General 
& \checkmark 
&  
& \checkmark 
& \checkmark 
& \checkmark \\

\midrule

AdvShield \cite{fan2025youadversarialshieldingvlmbased} 
& General 
& VLM
& 
& 
& 
& \checkmark
&
\\

HR-VISP \cite{abdulaziz_evaluation_2025} 
& \shortstack{PII, Social relationship} 
& CV models 
&  
& 
& \checkmark 
&  
&  \\

Imago Obscura \cite{monteiro_imago_2025} 
& General visual cues 
& Human 
& \checkmark 
& \checkmark 
& \checkmark 
& \checkmark 
&  \\

\midrule

\name{} (Ours)
& \shortstack{General\\(Geo, PII, Social relationship)} 
& \shortstack{General\\(Human, VLM, CV)} 
& \checkmark 
& \checkmark 
& \checkmark 
& \checkmark 
& \checkmark \\

\bottomrule
\end{tabular}%
}
\end{table*}

\section{Related Work}\label{sec:related_work}

\subsection{Visual Privacy Leakage Beyond Explicit Visible Cues}

Although image sharing is ubiquitous, users often remain unaware of the privacy risks their images expose~\cite{charnsethikul_navigating_2025,3489212.3489302}.
Such risks extend beyond explicit identifiers: multiple weak contextual cues may jointly reveal sensitive attributes through \emph{compositional inference}~\cite{tsaprazlis_rethinking_2026}.
VLMs amplify this threat by reasoning over contextual evidence beyond directly visible markers~\cite{tsaprazlis_assessing_2025}, while human-centered studies show that privacy perception similarly depends on co-presence, capture context, and higher-level interpretation~\cite{xu_dipa2_2023}.
Consequently, traditional classifiers that assign public/private labels using predefined scenes or objects cannot adequately capture such leakage~\cite{xompero_learning_2025}.
\name{} therefore analyzes and mitigates privacy risks arising beyond explicit visible cues, focusing on the following representative threats while remaining extensible to others.

\noindent \textbf{Visual geolocation inference attacks.}
Visual geolocation attacks infer an image's capture location from its visual content.
Early work frames this task as classification or retrieval against geo-tagged databases or discretized geographic grids~\cite{schubert2023visual}.
Recent VLM-based attacks instead reason over heterogeneous cues such as architecture, vehicles, spatial structure, and cultural elements, with chain-of-thought and multi-step reasoning further improving inference~\cite{liu2025mission,li_recognition_2025,yerramilli_geochain_2025,luo_doxing_2025}.
These advances enable localization from the kilometer scale down to individual blocks or streets.

\noindent \textbf{Personal identity and social relationship reasoning.}
Identity and relationship leakage can arise from interpersonal configurations, body language, surrounding people, and scene context~\cite{244030,285429}.
Relational attributes therefore emerge from person-pair interactions and contextual semantics rather than isolated objects~\cite{li_dual-glance_2017}.
Recent systems strengthen these inferences by converting visual content into social narratives~\cite{li_socialgpt_2024} or combining evidence within and across personal images to infer sensitive, abstract user attributes~\cite{liu_eye_2025}.

Overall, these studies motivate visual privacy defenses that protect not only explicit identifiers, but also sensitive inferences derived through contextual reasoning.

\subsection{Defenses Against Visual Privacy Threats}
Tab.~\ref{tab:comparison} summarizes representative defenses against
visual privacy threats, organized by their protection scope and control
mechanisms.

\noindent\textbf{Defense against visual geolocation inference attacks.}
Prior work primarily protects against specialized CV models that infer
geolocation from images.
ADGAN~\cite{xiong_adgan_2021}, LocGuard~\cite{ma_locguard_2024}, and
MM-PGD~\cite{le_rethinking_2022} use adversarial image modifications to
reduce geolocation leakage against CV-based attackers.
However, these methods are often optimized for particular model families
or feature spaces~\cite{308156,298176}, limiting their transfer to
semantic attackers such as VLMs.
Recent methods begin to address VLM-based attacks.
GeoShield~\cite{liu_geoshield_2026} perturbs geo-semantic features, while
ReasonBreak~\cite{zhang2026disruptinghierarchicalreasoningadversarial}
disrupts hierarchical geographic reasoning chains in multimodal reasoning
models.
Nevertheless, these defenses primarily optimize attack failure and do not
make the evidence supporting an inference individually controllable under
a user-selected protection boundary.

\noindent\textbf{Defense against PII and general privacy leakage.}
Prior work protects explicit PII and localized privacy cues through
redaction, selective obfuscation, and task-aware
sanitization~\cite{orekondy_connecting_2018,shu_cardea_2018,
venugopalan_aragorn_2023,285485,307742}.
However, predefined sensitive categories are insufficient when multiple
benign cues jointly reveal sensitive attributes through compositional
reasoning~\cite{abdulaziz_evaluation_2025,
tsaprazlis_rethinking_2026}.
Recent systems broaden this scope: CAIAMAR uses multi-agent reasoning and
iterative auditing to anonymize context-dependent PII, while Imago Obscura
surfaces contextual privacy risks and supports user-guided
obfuscation~\cite{aufschlager_towards_2026,monteiro_imago_2025}.
Other generative defenses remain specialized to facial
identity~\cite{le_styleadv_2024,le_diffprivate_2025}.

\noindent\textbf{User-steerable privacy protection.}
Existing systems support context-dependent policies or user control over
detected risks, regions, and editing operations~\cite{shu_cardea_2018,
venugopalan_aragorn_2023,monteiro_imago_2025}.
However, they do not treat the permitted specificity of the sensitive
inference itself---such as country versus city versus street-level
location---as the protection target.
In contrast, \name{} uses a user-selected inference granularity to
determine which supporting clues are proposed for removal, permits
clue-level accept-or-reject decisions, and verifies residual leakage
against the same boundary.

\begin{figure*}[h]
    \centering

    \includegraphics[width=\textwidth]{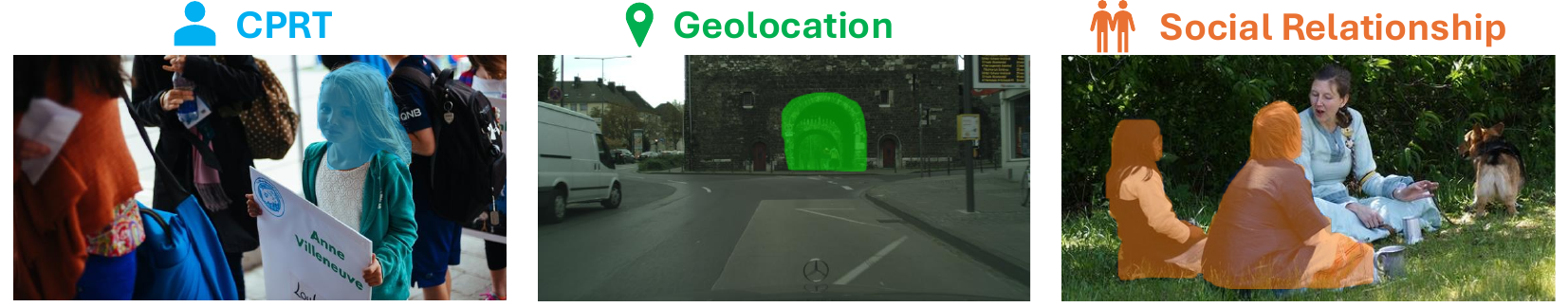}

    \vspace{0.8em}

    \footnotesize
    \begin{tabularx}{\textwidth}
    {p{0.04\textwidth} p{0.16\textwidth} X X X}
    \toprule
    & \textbf{Field} 
    & \textbf{CPRT} 
    & \textbf{Geolocation} 
    & \textbf{Social Relationship} \\
    \midrule

    \multirow{2}{*}{$l_i$}
    & \textbf{Granularity}
    & Unique identifier
    & City / Regional level
    & Intimate / non-intimate \\
    \addlinespace[2pt]
    
    & \textbf{Attribute}
    & Biometrics
    & Geographic location
    & Friends\\

    \midrule

    \multirow{2}{*}{$c_i$}
    & \textbf{Clue description}
    & Front-facing face of a young girl
    & Stone gate with a round arch
    & Two individuals sitting together \\
    \addlinespace[2pt]
    
    & \textbf{Target concept}
    & Face
    & Gate
    & Co-presence / interaction \\

    \midrule

    \multirow{3}{*}{$\pi_i$}
    & \textbf{Privacy irrelevant cues}
    & Green clothes
    & Rounded arch, Aged stone
    & Background scene \\
    \addlinespace[2pt]

    & \textbf{Privacy relevant cues}
    & Visible eyes, nose, mouth
    & Medieval style, Ancient gate
    & Posture, Proximity, Shared activity \\
    \addlinespace[2pt]

    & \textbf{Reasoning}
    & Identifiable frontal face of a child
    & Distinct medieval gate appearance matching Bonn Sterntor
    & Shared picnic context suggests close relationship \\

    \bottomrule
    \end{tabularx}

    \caption{
    Example clue visualization and its corresponding structured clue representation.
    The top row shows localized visual evidence from three privacy domains, and the table below maps each highlighted clue to its tuple components $(l_i, c_i, \pi_i)$. Images are from CPRT~\cite{tsaprazlis_rethinking_2026}, Cityscapes~\cite{cordts2016cityscapes}, and PISC~\cite{li_dual-glance_2017} dataset.
    }
    \Description{Three examples illustrate the structured privacy-clue representation. The CPRT example shows a person whose face is highlighted as privacy-relevant evidence; its structured fields identify a unique-identifier risk, biometrics as the attribute, the face as the target concept, facial features as privacy-relevant cues, and clothing as privacy-irrelevant content. The geolocation example highlights a stone gate whose architectural characteristics support a city-level location inference. The social-relationship example shows two people sitting together, where posture, proximity, and shared activity support an inferred friendship while the background scene is treated as irrelevant. A table below the images maps each example to its granularity, attribute, clue description, target concept, relevant and irrelevant cues, and reasoning.
    }
    \label{fig:clue_tuple_examples}
\end{figure*}

\section{Threat Model and Problem Definition}

\subsection{Attack Models}\label{sec:attackmodels}
We follow the visual privacy attack setting in~\cite{zhao_visual_2023}, where a victim captures a photo and shares it through a public or semi-public channel, such as social media, cloud service, online forums, or large messaging groups.
After the image is shared, a low-effort, non-expert adversary can obtain the image and use off-the-shelf inference tools, including OCR systems, object detectors, face or scene recognizers, search engines, public map services, or advanced VLMs, to infer sensitive information about the victim.
We assume the adversary only has access to the shared image at its platform-provided resolution and quality, without privileged access to private metadata, EXIF data, or other victim-specific context beyond what can be inferred from the image and public information.

Unlike traditional visual privacy attacks that primarily rely on explicit identifiers, we focus on privacy threats that arise from \emph{contextual and compositional visual reasoning}.
Formally, given a shared image $x$, the adversary aims to produce a sensitive inference $y$ about a victim or scene-associated individual by reasoning over a set of visible cues $\mathcal{E}(x)$.

\noindent \emph{(1) Compositional Privacy Risk Taxonomy (CPRT) Attack}: We consider the potential privacy threats by following the CPRT framework proposed by~\cite{tsaprazlis_rethinking_2026}, which models visual privacy as a graded and compositional severity-assessment problem.

Besides, we consider two specific representative privacy threats requiring detailed reasoning:

\noindent \emph{(2) Geolocation inference attacks.}
The adversary aims to infer where the image was captured, i.e., geographic location (precise latitude and longitude).
Such inference may rely on signage fragments, architectural styles, storefronts, or other region-specific visual evidence.
Following recent work on VLM-based geolocation attacks~\cite{liu2025mission}, we focus on practical zero-shot semantic reasoning attacks, where the adversary queries publicly accessible VLMs or open-source models to derive and refine location hypotheses from a single shared image.

\noindent \emph{(3) Social relationship inference attack.}
The adversary aims to infer the social relationships among people appearing in the image.
Following the visual social relationship recognition setting introduced by~\cite{li_dual-glance_2017}, we formulate the attack as a pairwise inference problem: given an image $I$ and two people $p_i$ and $p_j$ localized in the image, the adversary predicts a relationship label $r_{ij}$ between them.

The detailed defense granularities against these attacks are discussed in the next section (Sec~\ref{sec:defense_model}).

\subsection{Defense Models}
\label{sec:defense_model}

\noindent \textbf{Defense setting.} \name is a \emph{user-side} defense framework designed for a \emph{black-box} and \emph{model-agnostic} setting.
\name is applied \emph{before sharing}: it edits the raw image into a defended image that is then uploaded.
We assume the \name pipeline is operated in a trusted execution environment (TEE), that will not itself leak information beyond the defended image.
\name does not rely on knowing which model the adversary uses and assumes completely no access to the attack model parameters.
We consider \emph{adaptive adversaries} who may know the existence and strategy of \name, but can only operate on the defended image.
As mentioned in the attack model, we do not aim to defend against (i) privileged platform adversaries, (ii) attackers with additional private context beyond the shared image, or (iii) privacy experts who perform substantial manual investigation using external knowledge.

\noindent \textbf{Privacy protection objective.}
Given an input image $x$ and a user-specified granularity preference $g$, \name outputs a defended image
$x'=\name(x,g)$ that reduces privacy leakage beyond the level the user is willing to reveal.
We define $g$ using a task-specific granularity schema: 

\noindent \emph{(1) Compositional inference defense:} Level~1 denotes unique identifiers; Level~2 denotes linkage-based identifiers; Level~3 denotes aggregation-based identifiers; and Level~4 denotes
benign contextual information \cite{tsaprazlis_rethinking_2026}.

\noindent \emph{(2) Geolocation inference defense:}
We assess geolocation privacy protection across five privacy levels,
defined by the localization accuracy: \emph{block}
($\leq 0.5$\,km), \emph{street} ($\leq 1$\,km), \emph{neighborhood}
($\leq 5$\,km), \emph{city} ($\leq 10$\,km), and \emph{regional}
($\leq 100$\,km).

\noindent \emph{(3) Social relationship inference defense:}
We evaluate social relationship protection at two levels of granularity:
a 3-way classification, including \emph{intimate},
\emph{non-intimate}, and \emph{no relationship}; and a 6-way
classification, including \emph{friends}, \emph{family}, \emph{couples},
\emph{professional}, \emph{commercial}, and \emph{no relationship}, ordered from higher to lower privacy severity.

Such granularity serves as the protection policy for three reasons.
First, it distinguishes weak contextual leakage from precise or high-severity
inference, providing the pipeline with a priority signal for deciding which
clues should be edited, weakened, or preserved.
Second, it provides an intuitive user-facing control interface: users specify
the level of inference they are willing to reveal, rather than tuning low-level
model thresholds or editing parameters.
Third, it makes the defense decision explainable by linking each edit to the
inferred risk level it aims to suppress.
\name{} does not guarantee perfect removal of all sensitive information.
Instead, it aims to suppress visual evidence that could support inferences more sensitive than the user's selected privacy granularity.

\noindent \textbf{Usability and trustworthiness objectives.}
In addition, \name{} aims to satisfy two usability and
trustworthiness goals.

First, \name{} should \emph{preserve visual fidelity and image utility}.
The defended image $x'$ should remain suitable for normal sharing, with minimal
visual artifacts, limited unnecessary semantic loss, and preservation of
non-sensitive content relevant to the user's sharing intent.

Second, \name{} should provide \emph{explainability}.
Rather than outputting only a defended image, \name{} highlights the regions,
objects, relationships, or contextual cues that contribute to specific privacy
risks and explains why they matter.
These explanations help users understand how sensitive inferences arise from
compositional visual evidence, inspect the privacy--utility trade-off, and
choose a protection level that matches their sharing preferences.

\section{Methodology}
\label{sec:Methodology}

\begin{figure}[htbp]
    \centering
    \includegraphics[width=1\linewidth]{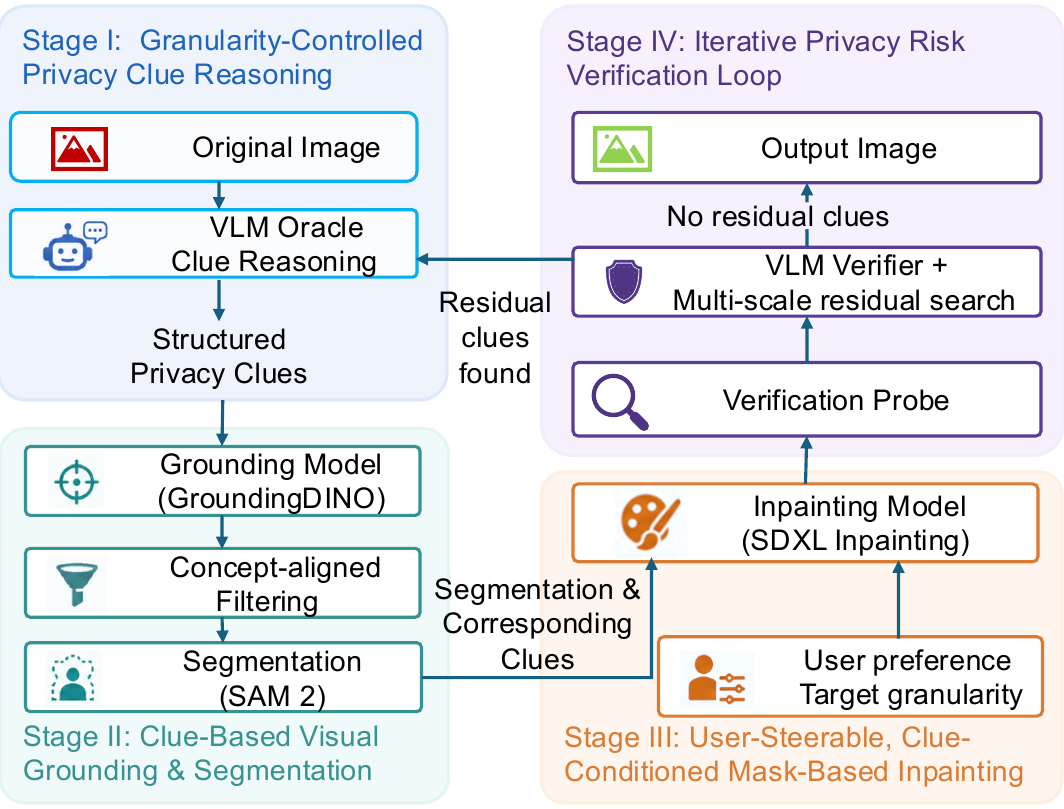}
    \caption{Overview of \name pipeline. It follows an iterative reasoning–localization–editing–verification workflow.}
    \label{fig:overview}
    \Description{Block diagram of the four-stage VeilLens pipeline. Stage I sends the original image to a VLM oracle that generates structured privacy clues under a user-selected target granularity. Stage II passes the clues through GroundingDINO, concept-aligned filtering, and SAM 2 segmentation to obtain localized masks. Stage III uses the masks, user preferences, and an SDXL inpainting model to edit privacy-relevant image regions. Stage IV evaluates the edited output with a VLM verifier and multi-scale residual search. If residual clues are found, they are returned to the grounding and editing stages for another iteration; otherwise, the protected output image is returned.
    }
\end{figure}

Given an input image $x$ and a user-selected protection granularity $g$, \name{} produces a defended image $x'$ through the iterative four-stage pipeline shown in Fig.~\ref{fig:overview}: privacy clue reasoning, visual grounding, user-steerable editing, and residual-risk verification.

\subsection{Stage I: Granularity-Controlled Privacy Clue Reasoning}
\label{subsec:clue-reasoning}

Given an input image $x$, Stage~I identifies visual clues that may support
privacy-sensitive inference.
Rather than detecting a predefined set of sensitive object categories, VeilLens
uses a trusted VLM oracle to produce a finite set of structured privacy clues:
\begin{equation}
F_{\text{VLM}}(x)
\;\mapsto\;
\mathcal{C}=\{ e_i = (\ell_i, c_i, \pi_i) \}_{i=1}^{k},
\end{equation}
Each clue $e_i$ denotes a concrete privacy-relevant visual cue: 

\noindent 
\emph{(1) Privacy risk category and granularity ($\ell_i$).}
It characterizes the type and granularity of privacy risk associated with clue $e_i$.
The granularity levels follow the domain-specific definitions in Sec.~\ref{sec:defense_model}.
$\ell_i$ makes explicit how each clue participates in a composite inference structure.

\noindent \emph{(2) Groundable clue description ($c_i$).}
It specifies the visual evidence associated with clue $e_i$ for Stage~II grounding (Sec.~\ref{subsec:visual-grounding}).
We represent $c_i$ using two components: a \emph{clue description} and a \emph{target concept}. The clue description is a RefCOCO-style referring expression~\cite{kazemzadeh-etal-2014-referitgame}: a concise noun phrase beginning with the target concept and augmented with visible modifiers (e.g., color, shape, material, size) and spatial relations.
The target concept corresponds to the leading noun phrase and serves as a semantic anchor for token-aligned candidate filtering (Sec.~\ref{subsec:visual-grounding}).

\noindent\emph{(3) Explainability and actionability ($\pi_i$).}
The field $\pi_i$ provides an actionable attribution for clue $e_i$ by separating \emph{privacy-relevant cues}, \emph{privacy-irrelevant cues}, and a brief \emph{reasoning} statement. The privacy-relevant elements specify what semantics should be suppressed, while the privacy-irrelevant cues describe visual properties that should be preserved when possible. The reasoning statement explains why these cues support the sensitive inference, helping users judge whether the proposed edit matches their own sharing intent before accepting or rejecting the intervention. This decomposition directly supports Stage~III (Sec~\ref{subsec:stage3}).

Fig.~\ref{fig:clue_tuple_examples} shows examples of $e_i$ across different privacy risks. 

We use a VLM oracle because privacy-relevant evidence is open-ended and 
heterogeneous, making it difficult to capture with predefined detectors. 
The oracle also produces natural-language rationales that support clue 
interpretation and downstream editing. 
\name treats the oracle as a replaceable component and assumes that 
it operates within the trusted defense boundary defined in 
Sec~\ref{sec:defense_model} where it may access $x$ but cannot 
disclose the unprotected image or its sensitive content.

\subsection{Stage II: Clue-Based Visual Grounding and Segmentation}
\label{subsec:visual-grounding}

\begin{figure*}[t]
    \centering
    \includegraphics[width=\textwidth]{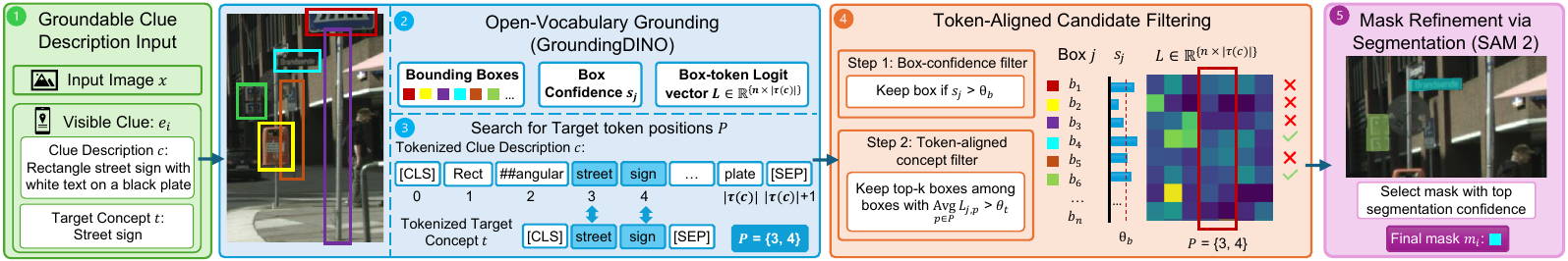}
    \caption{Visual grounding and segmentation pipeline. The target concept is located within the tokenized clue description to obtain the relevant token positions $P$. GroundingDINO's bounding-box logit matrix is then filtered by box confidence and concept-specific token activation, retaining boxes that correspond to the intended concept before segmentation refinement.}
    \label{fig:token}
    \Description{Detailed grounding pipeline for a clue describing a rectangular street sign with white text on a black plate and the target concept "street sign." The clue description and target concept are separately tokenized, and the positions corresponding to the target-concept tokens are identified. GroundingDINO produces candidate bounding boxes, box-confidence scores, and a box-by-token alignment-logit matrix. Candidate boxes are first filtered by overall confidence and then by activation on the target-concept token positions. The highest-ranked surviving boxes are passed to SAM 2, which generates segmentation masks, and the mask with the highest segmentation confidence is selected as the final localized clue region.
    }
\end{figure*}

Second, \name grounds the structured clue set 
$\mathcal{C}=\{e_i=(\ell_i,c_i,\pi_i)\}_{i=1}^{k}$ produced in Stage~I
into editable image regions, 
as shown in Fig.~\ref{fig:token}.

\noindent \textbf{\textcircled{1} Groundable Clue Description Input.}
A central requirement for Stage~II is that the grounding query must be both semantically
specific and spatially localizable.
Therefore, as introduced in Stage~I, \name represents each clue description $c_i$ as a
RefCOCO-style referring expression~\cite{kazemzadeh-etal-2014-referitgame}, which provides a concrete instance-level query suitable for open-vocabulary
grounding, while preserving enough visual specificity to distinguish among similar
objects.

To validate this design choice, we compare RefCOCO-style expressions against two
representative alternatives for converting VLM-generated clues into grounding queries.
\emph{Free-form descriptions} directly use the VLM-generated explanation of a
privacy clue as the grounding query.
\emph{Categorical labels} reduce each clue to its primary object, scene, or privacy-attribute noun, such as \emph{license plate} and \emph{street sign}. This representation follows the attribute-style
formulation used in visual privacy datasets such as VISPR \cite{abdulaziz_evaluation_2025}.
For each generated clue, we run GroundingDINO~\cite{liu_grounding_2024} to
produce candidate bounding boxes from the corresponding grounding query.
A human annotator then inspects the top-1 confidence bounding box and counts the
result as successful only if the box tightly covers the intended target clue.
As shown in Tab.~\ref{tab:refcoco_prompt_examples}, RefCOCO-style expressions
achieve the highest grounding success rate.
It suggests that RefCOCO-style expressions provide a useful middle
ground: they are concise enough to be grounded reliably, while retaining
modifiers and spatial context needed to localize the intended instance.
Thus, we use RefCOCO-style expressions as our default representation of $c_i$.

\noindent \textbf{\textcircled{2} Open-Vocabulary Grounding.}
Given an image $x$ and a clue description $c_i$, we first apply an open-vocabulary grounding operator $\mathcal{G}$:
\begin{equation}
\mathcal{G}(x,c_i) \;\mapsto\; (B_i, L_i, S_i)
\end{equation}
where $B_i=\{b_{ij}\}_{j=1}^{n}$ denotes the $n$ candidate bounding boxes produced by $\mathcal{G}$,
$L_i\in\mathbb{R}^{n\times |\tau(c_i)|}$ is the box-token alignment logit matrix
over the tokenized clue $\tau(c_i)$, and
$S_i=\{s_{ij}\}_{j=1}^{n}\in\mathbb{R}^{n}$ contains the grounding confidence scores,
with $s_{ij}$ corresponding to candidate box $b_{ij}$.
We instantiate $\mathcal{G}$ using GroundingDINO~\cite{liu_grounding_2024}, an open-sourced text-conditioned open-vocabulary detector.

\noindent \textbf{\textcircled{3} Targeted Token Position Search.}
Although open-vocabulary grounding provides candidate boxes, the highest-confidence box
does not always correspond to the intended clue.
In particular, grounding can be activated by incidental modifiers, nearby objects, or
context words in the clue phrase rather than by the target object itself.
To improve semantic faithfulness, \name refines GroundingDINO candidates using the
token-level alignment scores produced by the detector.
We tokenize both $c_i$ and $t_i$ 
using the tokenizer in $\mathcal{G}$ and identify the set of
token positions $P_i(t_i,c_i)$ in $\tau(c_i)$ that correspond to $t_i$ by matching token IDs
between the target concept and the object description.

\noindent \textbf{\textcircled{4} Token-Aligned Candidate Filtering.}
We keep boxes that pass both confidence and target-token thresholds:
\begin{equation}
\mathcal{J}_i = \big\{j : s_{ij} > \theta_b \; \bigwedge \;
   \tfrac{1}{|P_i|}\!\sum_{p\in P_i} L_{i,j,p} > \theta_t\big\},
\end{equation}
If $P_i=\emptyset$ or the token-aligned filter yields $\mathcal{J}_i=\emptyset$, we instead use the confidence-filtered set
$\mathcal{J}_i=\{j:s_{ij}>\theta_b\}$.
We rank the remaining candidates and pass the top two boxes to SAM~2.

\noindent \textbf{\textcircled{5} Mask Refinement via Segmentation.}
The retained boxes approximately localize the clue but are too coarse for localized editing (Sec.~\ref{subsec:stage3}).
For each retained box $b_{ij}$, SAM~2~\cite{ravi2024sam2} produces pixel-level mask candidates with corresponding segmentation-confidence scores.
Across the mask candidates generated from the retained boxes, \name{} selects the highest-confidence mask $m_i$ as the final grounded region for clue $e_i$.

\begin{table}[t]
\centering
\caption{Comparison of clue-description formats used in Stage~I. The grounding success rate measures the fraction of generated clues with correctly produced bounding boxes.}
\label{tab:refcoco_prompt_examples}
\footnotesize
\setlength{\tabcolsep}{5pt}
\renewcommand{\arraystretch}{1.12}

\begin{tabularx}{\linewidth}{@{}
>{\raggedright\arraybackslash}p{0.24\linewidth}
>{\raggedright\arraybackslash}X
c
@{}}
\toprule
\textbf{Prompt format}
&
\textbf{Example output}
&
\makecell[c]{\textbf{Grounding} \\   \textbf{success rate}} \\
\midrule

Free-form\newline description
&
The medieval-looking entrance structure suggests a historic European city center.
&
0.76 \\
\addlinespace[0.7em]

Categorical label
&
Sign; Building.
&
0.64 \\
\addlinespace[0.7em]

RefCOCO-style\newline expression
&
City gate with dark stone arch; Yellow stop sign beside road.
&
0.89 \\

\bottomrule
\end{tabularx}

\end{table}

\subsection{Stage III: User-Steerable, Clue-Conditioned Mask-Based Generative Inpainting}
\label{subsec:stage3}

Given the structured clue set
$\mathcal{C}=\{e_i=(\ell_i,c_i,\pi_i)\}_{i=1}^{k}$ from Stage~I
and the corresponding masks $\{m_i\}_{i=1}^{k}$ from Stage~II,
Stage~III selects which clues to protect and applies localized
generative inpainting to their grounded regions.

\noindent \textbf{User-steerable privacy editing.}
Let $\ell_i\preceq g$ indicate that the risk level of clue $e_i$
falls within the task-specific protection boundary selected by
the user.
Stage~III constructs the candidate clue set and the user-accepted
edit set as
\begin{equation}
\mathcal{C}_g
=
\{e_i\in\mathcal{C}:\ell_i\preceq g\},
\qquad
\mathcal{E}_{g,\mathbf{a}}
=
\{e_i\in\mathcal{C}_g:a_i=1\},
\end{equation}
where $a_i\in\{0,1\}$ indicates whether the user accepts the
proposed intervention for clue $e_i$.
For example, city-level geolocation protection includes clues
supporting city-or-finer inference in $\mathcal{C}_g$, while
clues revealing only coarser regional information are retained
by default.
For each candidate, \name{} presents its risk annotation $\ell_i$,
rationale $\pi_i$, and localized mask $m_i$, allowing the user
to inspect the affected visual content before accepting or rejecting
the edit.

\noindent \textbf{Clue-conditioned localized editing.}
For each accepted clue $e_i\in\mathcal{E}_{g,\mathbf{a}}$,
\name{} applies localized generative inpainting to its corresponding
mask:
\begin{equation}
x'
=
\mathcal{I}
\bigl(x,m_i,q(x),p_i^{+},p_i^{-}\bigr),
\end{equation}
where $\mathcal{I}$ denotes a pretrained diffusion-based inpainting
model, $x$ is the current image, and $x'$ is the result after editing
clue $e_i$.
The structural condition $q(x)$ is a monocular depth map estimated
using a pretrained depth-prediction model%
~\cite{ranftl2020midas,ranftl2021visiontransformersdenseprediction}
and injected through ControlNet~\cite{zhang2023controlnet} to preserve
the surrounding scene geometry and object layout.
The privacy-irrelevant and privacy-relevant cues in $\pi_i$ form the
positive prompt $p_i^{+}$ and negative prompt $p_i^{-}$, respectively,
directing the model to preserve benign visual properties while
suppressing sensitive semantics.
When multiple clues are accepted, \name{} applies the edits
sequentially, using each edited image as the input to the next
operation.
After all accepted clues have been processed, the resulting image
is passed to Stage~IV for residual-risk verification.

\noindent \textbf{Global photo editing and risk feedback.}
In addition, \name also supports global photo edits that users may apply for aesthetic or sharing purposes.
Such edits include common photographic adjustments and filters that modify the entire image rather than a specific mask.
We found that some global effects may reduce residual privacy risk by weakening fine-grained visual cues used by VLM-based attackers.
In particular, film grain is a promising candidate because it is a familiar photographic effect and introduces tunable stochastic texture that can reduce the visibility of subtle cues while maintaining a natural image appearance.
This allows users to combine localized inpainting with ordinary photo editing while maintaining visibility into the current privacy risk of the shared image.

\subsection{Stage IV: Iterative Privacy Risk Verification Loop}
\label{subsec:stage4}

A single editing round may suppress the most salient clues while
leaving weaker evidence that still supports an inference beyond the
user-selected protection granularity $g$.
\name{} therefore alternates full-image residual-risk verification
with targeted clue discovery and re-editing.

\noindent \textbf{Full-image verification and residual-clue discovery.}
Let $x^{(t)}$ denote the image after editing round $t$, and let
$M^{(t)}$ denote the cumulative union of masks for clues already
presented through round $t$.
The verifier first evaluates the complete, unmasked image $x^{(t)}$.
If it detects no inference beyond $g$, the loop terminates.
Otherwise, \name{} constructs a discovery probe by masking the
previously handled clue regions:
\begin{equation}
\tilde{x}^{(t)}
=
\operatorname{Mask}
\bigl(x^{(t)},M^{(t)}\bigr).
\end{equation}
The probe is used only to discover additional clues and not to
determine whether protection has succeeded.
\name{} re-runs the Stage~I clue-reasoning module on
$\tilde{x}^{(t)}$, preventing repeated discovery of the same dominant
evidence and directing the search toward previously unaddressed clues.
Newly discovered clues within the selected protection boundary are
grounded and presented for editing through the same Stage~II--III
pipeline, producing the next edited image $x^{(t+1)}$.

\noindent \textbf{Multi-scale residual search and termination.}
If the full-image verifier still detects an inference beyond $g$ but
the global discovery probe yields no actionable clue, \name{} searches
over overlapping patches of $\tilde{x}^{(t)}$ at multiple scales.
Each patch is evaluated in the context of the residual full-image
inference, and only locally visible evidence that supports that
inference is retained for grounding and editing.
The loop terminates successfully when the full-image verifier detects
no inference beyond $g$.
If the iteration budget is exhausted or no additional actionable clue
can be found, \name{} returns the latest image and reports that the
requested protection level remains unverified.

\section{Implementation and Default Setup}
\label{sec:impl}
We implement \name{} using commercial off-the-shelf VLM APIs and open-source vision models. All experiments run on an NVIDIA RTX PRO 6000 GPU. All images are evaluated at original resolution.

\noindent \textbf{Reasoning and verification VLM calls.}
All defense-side VLM calls return structured outputs following task-specific schemas.
We use GPT-5 as the default trusted VLM oracle for Stage~I privacy-clue reasoning and Stage~IV residual-risk verification.
This oracle is used only inside the defense pipeline and is kept separate from the adversary VLMs used for evaluation.
The VLM attackers used for evaluation are described as attacker baselines in Sec.~\ref{sec:evaluation-design}.
Keeping the defense oracle distinct from the evaluation attackers avoids overestimating protection due to model-specific alignment between the defender and the attacker, and is consistent with our model-agnostic defense design.
Importantly, GPT-5 is only the default instantiation of the trusted oracle; \name{} treats the VLM oracle as a replaceable component; App.~\ref{app:oracle} evaluates a locally served open-weight oracle.

\noindent \textbf{Grounding and Editing.}
For each clue, the grounding stage queries GroundingDINO~\cite{liu_grounding_2024} (\texttt{SwinT-OGC}) using its default 900 object queries (candidate boxes), applies a box threshold of $0.30$, retains the top two candidates, and refines them into pixel-level masks via SAM 2.1~\cite{ravi2024sam2}; %
Editing backend is SDXL 1.0~\cite{podell2023sdxl} with a depth-ControlNet (\texttt{controlnet-depth-sdxl-1.0-small})~\cite{zhang2023controlnet} conditioned on MiDaS \texttt{DPT-Hybrid}~\cite{ranftl2020midas} depth maps ($50$ denoising steps, guidance $7.5$, ControlNet scale $0.7$, strength $0.8$);
The default pipeline runs verifier at most twice: Round~1 on the full image; Round~2 optionally with multi-scale residual search using overlapping $768\times768$ crops at $10\%$ overlap. The full configuration, runtime, and prompts are given in App.~\ref{app:impl} and App.~\ref{app:prompts}.

\begin{table*}[t]
\centering
\small
\caption{Defense baselines used in our evaluation.}
\label{tab:methods}
\resizebox{\textwidth}{!}{%
\begin{tabular}{lll}
\toprule
\textbf{Name} & \textbf{Design axis} & \textbf{Core mechanism} \\
\midrule
\textsc{raw}
  & ---
  & Image without defense\\
\textsc{rand-noise}
  & Non-semantic global
  & Replace 10\% randomly selected pixels with values from a $51{\times}51$ Gaussian-blurred image \\
\textsc{ocr-blur}
  & Semantic text obfuscation
  & EAST detection~\cite{zhou2017east} + $31{\times}31$ Gaussian blur on text regions \\
\textsc{mm-pgd~\cite{le_rethinking_2022}}
  & Adversarial perturbation
  & PGD~\cite{madry2018towards} attack against a Places365 ensemble, $\ell_\infty$ $\epsilon{=}0.5$, $T{=}20$ \\
\textsc{geoshield~\cite{liu_geoshield_2026}}
  & CLIP-ensemble adversarial (geo-only)
  & FGSM against CLIP surrogates with geo-semantic loss~\cite{liu_geoshield_2026}, square-tiled \\
\textsc{gpt-image-2~\cite{openai_gpt_image_2}}
  & General-purpose generative editing
  & One-shot instruction-conditioned editing using the target privacy objective \\
\textbf{\name{}}
  & VLM-guided localized editing
  & Reasoning-guided clue grounding, inpainting, and verification\\
\bottomrule
\end{tabular}
}
\end{table*}

\section{Experimental Setup}
\label{sec:evaluation-design}

Our evaluation covers the three categories of privacy risk introduced in our threat model (Sec.~\ref{sec:defense_model}): \emph{compositional inference}, \emph{geolocation inference}, and \emph{social-relationship inference}.
Across tasks, \name{} follows the same pipeline, with task-specific schemas, metrics, and intervention policies.

\noindent\textbf{General Compositional Privacy Risk.}

\noindent\emph{Metric.}
We use the Continuous Privacy Risk Score from CPRT~\cite{tsaprazlis_rethinking_2026}, which maps privacy-relevant attributes to a severity value in $[0,1]$.
Unlike binary privacy labels, this score captures both explicit identifiers and weaker compositional cues.
We report both the post-defense score ($\downarrow$ better) and its reduction relative to the undefended image ($\uparrow$ better). 
We set the compositional protection target to CPRT Level 2, suppressing both level 1 and 2 identifiers.

\noindent\emph{Dataset.}
We evaluate on $500$ randomly sampled images from the CPRT dataset~\cite{tsaprazlis_rethinking_2026}, which reorganizes VISPR annotations~\cite{orekondy_towards_2017} into graded severity levels and continuous risk scores.
The subset has a mean severity of $0.48$, spanning both low- and high-risk images.

\noindent\textbf{Geolocation Privacy Risk.}

\noindent\emph{Metric.}
We measure geodesic error ($\uparrow$) as the Haversine distance between
the predicted and ground-truth coordinates. We evaluate attack success
at the granularity thresholds defined in Sec.~\ref{sec:defense_model}:
\emph{block} ($0.5$\,km), \emph{street} ($1$\,km),
\emph{neighborhood} ($5$\,km), \emph{city} ($10$\,km), and
\emph{regional} ($100$\,km). At each threshold, attack success rate is
the fraction of predictions within the corresponding distance
($\downarrow$). We use city-level protection, treating predictions within
$10$\,km as privacy failures.

\noindent\emph{Dataset.}
We evaluate on $500$ images from \textsc{Cityscapes}~\cite{cordts2016cityscapes}, sampled uniformly across $18$ European cities (App.~\ref{app:datasets}).
Ground-truth GPS coordinates are extracted from the provided vehicle metadata.
Cityscapes is challenging for the defender because its dashcam images contain subtle, distributed urban cues. For cross-dataset robustness, we also build a $200$-image \textsc{Mapillary} holdout set~\cite{mapillary_api}.
We sample $20$ global cities with population at least $300{,}000$ and collect $10$ non-panoramic street-level images near each city.
Images are sampled uniformly within a fixed radius around each city center, subject to Mapillary availability and the non-panoramic constraint.
Each image uses its own Mapillary GPS coordinate as ground truth.
We report \textsc{Mapillary} separately because it differs from \textsc{Cityscapes} in capture device, viewpoint, geography, and image distribution.

\noindent\emph{Real-world privacy relevance.} Although Cityscapes was not collected specifically for privacy research, its street-level imagery contains the same distributed urban cues that VLMs exploit when inferring the location of ordinary shared photographs. We therefore use it as a controlled benchmark for visual location leakage, and separately evaluate on geographically diverse Mapillary imagery to test transfer beyond the Cityscapes capture distribution.

\noindent\textbf{Social-Relationship Privacy Risk.}

\noindent \emph{Metric.}
We report two quantities.
(i) \emph{Relationship Attack Accuracy} ($\downarrow$ better): the adversary VLM's relationship-classification accuracy on annotated person pairs, measured before and after defense. We report this under both the 3-way and 6-way PISC labels (Sec.~\ref{sec:defense_model}).
(ii) \emph{Attack Reduction} ($\uparrow$ better): the relative decrease in attack accuracy after defense, computed as $(\mathrm{Acc}_{raw}-\mathrm{Acc}_{def})/\mathrm{Acc}_{raw}$.
We set the social-relationship target to no relationship, suppressing evidence for both intimate and non-intimate relationship inference. 

\noindent\emph{Intervention policy.}
For social-relationship privacy, we use dataset-provided person boxes and IDs to isolate relationship reasoning from person discovery. During evaluation, \name{} reasons about the specified subject pair, and once a relationship-bearing pair is selected, we directly anonymize the face regions within the corresponding boxes. This does not reduce generality, since subject localization can be supplied by standard person detectors.

\noindent\emph{Dataset.} We use PISC~\cite{li_dual-glance_2017}, which provides person bounding boxes and pairwise relationship labels.
We randomly sample $454$ images with $1{,}109$ annotated pairs across six categories: friends ($307$), family-members ($109$), couple ($37$), professional ($397$), commercial ($29$), and no-relationship ($230$).
During preprocessing, we overlay dataset-provided boxes and IDs so VLM predictions can be matched to subject pair.

\noindent\textbf{Perceptual Quality Metrics.}
To evaluate utility, we use four perceptual metrics: \emph{SSIM}~\cite{wang2004image} for structural similarity ($\uparrow$), \emph{LPIPS}~\cite{zhang2018lpips} for learned perceptual distance ($\downarrow$), \emph{CLIP-IQA}~\cite{wang2023clipiqa} for no-reference quality ($\uparrow$), and \emph{LIQE}~\cite{zhang2023liqe} for learned image quality evaluation ($\uparrow$).

\noindent\textbf{Attacker Baselines.}
We evaluate against four frontier adversary VLMs: \texttt{gemini-3-flash-preview}~\cite{google2025gemini3flash}, \texttt{gemini-3.1-pro-preview} \cite{google2026gemini31pro}, \texttt{gemini-2.5-flash}~\cite{geminiteam2025gemini25}, and \texttt{gpt-5.2}~\cite{openai2025gpt52}.
These models span multiple providers and capability tiers.
The attacker set includes models different from, and in some cases newer than, the defense oracle to reduce model-specific evaluation bias.

\noindent\textbf{Defense Baselines.}
We compare \name{} with a no-defense baseline, four representative privacy-defense baselines, and a frontier general-purpose baseline, as summarized in Tab.~\ref{tab:methods} (settings in App.~\ref{app:baselines}). The privacy-defense baselines cover non-semantic perturbation, text obfuscation, adversarial perturbation, and CLIP-based geo-semantic perturbation, with geolocation-specific methods evaluated only on the geolocation task. In addition, we include GPT-Image-2 as a strong general-purpose generative editing baseline, which directly edits the input image according to the same task-level privacy objective provided to \name{}, without access to \name{}'s discovered clues, localized masks, or verification feedback.

\noindent\textbf{Anti-Caching Mechanisms.}
\label{subsec:anti-caching} Repeated API queries may be affected by response, prefix, or key-value caching in modern VLM serving stacks, potentially producing correlated outputs or timing artifacts~\cite{song2024earlybird, gao2025iknow}.
To reduce such artifacts, we prepend each request with a UUIDv4 nonce, formatted as \texttt{Request-ID: <uuid>}, and instruct the model to ignore it.
\begin{figure*}[t]
\centering

\captionof{table}{\textbf{Top:} Geolocation attack success rates on the 500-image benchmark. Lower is better. \textbf{Bottom:} survival curves; higher curves indicate stronger protection. Bold and underline denote the best and second-best results.}
\label{tab:combined_evalcentric}
\setlength{\tabcolsep}{1.3pt}
\renewcommand{\arraystretch}{0.95}
\resizebox{\textwidth}{!}{%
\begin{tabular}{lrrrrrrrrrrrrrrrrrrrr}
\toprule
& \multicolumn{5}{c}{\textbf{Gemini-3.1-Pro-Preview}}
& \multicolumn{5}{c}{\textbf{Gemini-3-Flash-Preview}}
& \multicolumn{5}{c}{\textbf{Gemini-2.5-Flash}}
& \multicolumn{5}{c}{\textbf{GPT-5.2}} \\
\cmidrule(lr){2-6}
\cmidrule(lr){7-11}
\cmidrule(lr){12-16}
\cmidrule(lr){17-21}
\textbf{Error(km)}
& \textbf{$\leq$0.5} & \textbf{$\leq$1} & \textbf{$\leq$5} & \textbf{$\leq$10} & \textbf{$\leq$100}
& \textbf{$\leq$0.5} & \textbf{$\leq$1} & \textbf{$\leq$5} & \textbf{$\leq$10} & \textbf{$\leq$100}
& \textbf{$\leq$0.5} & \textbf{$\leq$1} & \textbf{$\leq$5} & \textbf{$\leq$10} & \textbf{$\leq$100}
& \textbf{$\leq$0.5} & \textbf{$\leq$1} & \textbf{$\leq$5} & \textbf{$\leq$10} & \textbf{$\leq$100} \\
\midrule
raw
& 40.2\% & 55.2\% & 73.6\% & 74.4\% & 87.2\%
& 41.6\% & 54.8\% & 70.4\% & 70.4\% & 86.4\%
& 15.8\% & 29.6\% & 45.0\% & 46.4\% & 69.2\%
& 11.4\% & 25.6\% & 39.2\% & 39.8\% & 61.6\% \\
\addlinespace[2pt]

rand-noise
& 28.0\% & 41.8\% & 57.2\% & 57.8\% & 74.6\%
& 27.4\% & 38.2\% & 51.8\% & 52.2\% & 74.2\%
& 11.8\% & 20.4\% & 31.2\% & 31.8\% & 57.2\%
& \underline{4.0\%} & 11.2\% & 19.2\% & 19.8\% & \underline{37.4\%} \\
\addlinespace[2pt]

ocr-blur
& 37.6\% & 50.2\% & 69.0\% & 70.0\% & 82.4\%
& 35.4\% & 46.4\% & 63.8\% & 64.6\% & 81.2\%
& 14.4\% & 28.2\% & 41.6\% & 42.8\% & 64.8\%
& 9.8\% & 21.4\% & 35.8\% & 36.2\% & 57.2\% \\
\addlinespace[2pt]

mm-pgd
& 30.4\% & 44.4\% & 60.0\% & 60.8\% & 75.4\%
& 29.2\% & 42.6\% & 57.2\% & 58.0\% & 78.8\%
& 14.6\% & 25.6\% & 38.0\% & 39.2\% & 64.8\%
& 8.8\% & 18.4\% & 30.6\% & 31.4\% & 52.2\% \\
\addlinespace[2pt]

geoshield
& 25.2\% & 36.8\% & 49.0\% & 49.8\% & 66.8\%
& 27.2\% & 37.4\% & 50.8\% & 51.6\% & \underline{67.4\%}
& 9.6\% & 17.4\% & 27.6\% & 28.4\% & \textbf{43.0\%}
& 6.4\% & 15.4\% & 24.4\% & 24.8\% & 38.8\% \\
\addlinespace[2pt]

GPT-Image-2
& \underline{17.4\%} & \underline{23.8\%} & \underline{37.8\%} & \underline{38.2\%} & \textbf{50.4\%}
& \textbf{14.2\%} & \underline{22.4\%} & \textbf{32.6\%} & \textbf{33.8\%} & \textbf{55.0\%}
& \textbf{7.6\%} & \underline{15.8\%} & \underline{26.6\%} & \underline{27.6\%} & 54.4\%
& 4.8\% & \underline{8.6\%} & \underline{16.6\%} & \underline{17.0\%} & 40.2\% \\

\midrule

\textbf{\name{}}
& \textbf{14.6\%} & \textbf{23.2\%} & \textbf{35.8\%} & \textbf{36.4\%} & \underline{52.8\%}
& \underline{14.4\%} & \textbf{22.2\%} & \underline{33.8\%} & \underline{34.0\%} & \textbf{55.0\%}
& \underline{8.2\%} & \textbf{14.8\%} & \textbf{21.2\%} & \textbf{21.8\%} & \underline{46.6\%}
& \textbf{1.2\%} & \textbf{4.8\%} & \textbf{10.8\%} & \textbf{11.0\%} & \textbf{26.2\%} \\
\bottomrule
\end{tabular}%
}

\vspace{0.7em}

\includegraphics[
        width=\textwidth
    ]{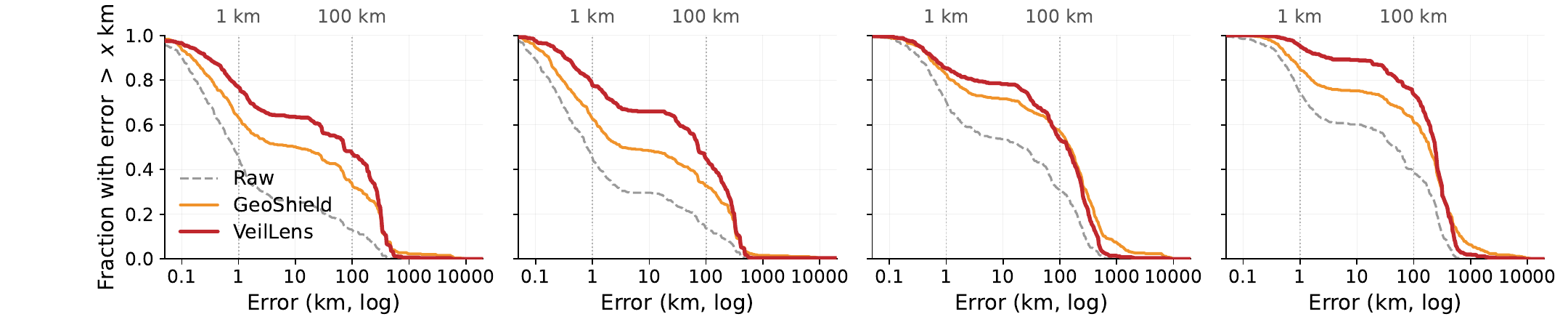}
    \label{fig:combined_results}

\end{figure*}

\section{Quantitative Evaluation Results}
\label{sec:results}

Next, we quantitatively evaluate \name{} against three representative privacy risks as discussed in the previous section.

\subsection{General Compositional Privacy Risk Results}

\begin{figure}[htbp]
    \centering
    \includegraphics[width=\linewidth]{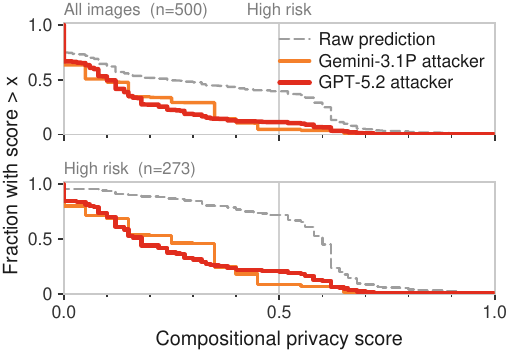} %
    \caption{CPRT privacy-risk score distributions before and after protection; lower is better. 
    Top: all images. Bottom: high-risk subset with ground-truth CPRT score $\geq 0.5$. 
    }
    \label{fig:cprt_risk_curves}
    \Description{Two privacy-risk survival plots compare raw images with VeilLens-protected images evaluated by GPT-5.2 and Gemini-3.1P. The horizontal axis is the continuous compositional privacy-risk score from zero to one, and the vertical axis is the fraction of images whose score exceeds each threshold. One plot covers all 500 images and the other covers the 273 high-risk images whose ground-truth CPRT score is at least 0.5. In both plots, the protected-image curves lie below the raw-image curve across most thresholds, indicating lower residual privacy risk after protection. The reduction is also visible on the high-risk subset.
    }
\end{figure}

Fig.~\ref{fig:cprt_risk_curves} reports CPRT privacy-risk distributions before and after protection under GPT-5.2 and Gemini-3.1P attackers.

\noindent \textbf{Raw images expose measurable compositional privacy risk.}
We evaluate general compositional privacy using the CPRT privacy risk score, where higher scores indicate stronger model-inferred privacy leakage.
On the raw images, the mean privacy risk score is $0.3138$, showing that many images contain privacy-relevant evidence beyond explicit identifiers.
This confirms the central motivation of \name{}: ordinary images can expose sensitive information through weak and compositional cues, rather than only through obvious objects such as faces, text, or license plates.

\noindent \textbf{\name{} substantially reduces model-inferred compositional privacy risk.}
Under GPT-5.2, the mean CPRT privacy risk score drops from $0.3138$ on raw images to $0.1611$ after protection, corresponding to a $48.7\%$ relative risk reduction.
Gemini-3.1P shows a similar trend, reducing the mean score from $0.3138$ to $0.1641$, a $47.7\%$ relative reduction.
This consistency across two frontier attackers suggests that the reduction is not specific to a single model family.
As shown in Fig.~\ref{fig:cprt_risk_curves}, both post-defense survival curves are consistently below the raw curve, indicating that \name{} shifts the full risk distribution toward lower privacy leakage rather than improving only a small subset of images.
The more step-like Gemini curve reflects Gemini-3.1P's coarser risk-score output granularity, not a smaller evaluation set or a substantively different trend.

\noindent \textbf{\name{} is especially effective on high-risk images.}
Among the $273$ high-concern images with ground-truth CPRT score $\geq 0.5$, the mean privacy risk score decreases from $0.5224$ to $0.2456$ under both GPT-5.2 and Gemini-3.1P.
This corresponds to a $53.0\%$ relative reduction, larger than the reduction on the full set.
These results show that \name{} reduces high-severity Level~1 and Level~2 leakage, rather than only improving already low-risk cases.

\noindent \textbf{\name{} achieves these reductions with localized rather than destructive edits.}
The median mask coverage is only $0.18\%$ of image pixels, and roughly half of the images are correctly left untouched.
The mean mask coverage of $14.3\%$ reflects a smaller subset of images requiring larger interventions.
This distribution supports the intended privacy--utility trade-off: \name{} applies protection selectively to reasoning-relevant regions instead of relying on blanket blurring, global degradation, or face-only anonymization.

\subsection{Geolocation Privacy Risk Results}\label{sec:eval::geo}
Tab.~\ref{tab:combined_evalcentric} reports geolocation attack success rates before and after each defense baseline across four VLM attackers and five distance thresholds.

\noindent \textbf{Across all evaluated attackers, \name{} substantially reduces fine-grained geolocation success.}
Undefended images expose substantial geolocation risk under recent strong VLMs.
At the most privacy-sensitive $0.5$\,km threshold, Gemini-3.1P achieves a $40.2\%$ success rate on raw images and Gemini-3F a $41.6\%$ rate, compared with $15.8\%$ for Gemini-2.5F and $11.4\%$ for GPT-5.2.
This suggests that more capable VLMs can localize a large fraction of images to near block-level precision, making geolocation leakage an increasingly severe threat.
\name{} substantially reduces this risk.
Under Gemini-3.1P, the strongest attacker we evaluate, \name{} reduces the raw $0.5$\,km success rate from $40.2\%$ to $14.6\%$, corresponding to a $2.75\times$ decrease.

\noindent \textbf{\name{} provides the strongest protection against the stron-gest attackers.}
For Gemini-3.1P at the $0.5$\,km threshold, the strongest geolocation-specific prior baseline is \textsc{GeoShield}~\cite{liu_geoshield_2026} at $25.2\%$.
\name{} further reduces success to $14.6\%$, a $1.73\times$ lower attack success rate than the strongest prior defense.

Under the Gemini-3.1-Pro attacker, the strongest attacker, GPT-Image-2 achieves $17.4\%$, while \name{} achieves $14.6\%$ for block level($\leq$ 0.5km). This indicates that \name{} outperforms GPT-Image-2 on the hardest scenario, while \name{} additionally provides explicit clue attribution, granularity control, and iterative verification.

\noindent \textbf{\name{} provides consistent protection across distance \newline thresholds and is especially effective against high-risk images.}
To assess whether \name{} improves geolocation privacy beyond a few selected thresholds, Tab.~\ref{tab:combined_evalcentric} reports error distributions and survival curves for \textsc{Raw}, \textsc{GeoShield}, and \name{}.
For threshold $x$, the y-axis of the survival curve denotes the fraction of images with geolocation error exceeding $x$, so higher curves indicate stronger protection.
Across all four evaluators, \name{} remains above both \textsc{Raw} and \textsc{GeoShield} over most of the threshold range, showing that the benefit reflects a broad shift in the full error distribution rather than improvement at only a few discrete cutoffs.

Compared with GPT-Image-2, \name{} reduces attack success rates across $13$ of $20$ evaluated attacker--threshold combinations, including all five thresholds under GPT-5.2 and four of five under Gemini-3.1P, while within two percentage points on five combinations.

The advantage is particularly clear in the privacy-sensitive regime below $10$\,km, where successful localization can reveal block-, street-, or district-level information.
For Gemini-2.5F, the improvement over \textsc{GeoShield} is more modest because the raw geolocation risk is already lower for this weaker attacker.

\noindent \textbf{\name{} generalizes across image datasets.}
We further evaluate \name{} on a $200$-image \textsc{Mapillary} dataset to test whether its geolocation protection transfers beyond the \textsc{Cityscapes} distribution.
Compared with \textsc{Cityscapes}, \textsc{Mapillary} contains more diverse geographies, camera sources, viewpoints, and scene compositions, making it a stronger test of dataset robustness.
\name{} substantially reduces fine-grained geolocation success on this set.
At the most privacy-sensitive $0.5$\,km threshold, Gemini-3F achieves a $14.5\%$ success rate on raw images, while \name{} reduces the success rate to $4.5\%$, corresponding to a $3.22\times$ decrease.
In comparison, \textsc{GeoShield} only reduces the success rate to $11.0\%$.
The same trend holds at coarser thresholds: \name{} reduces $1$\,km success from $21.0\%$ to $14.0\%$ and $10$\,km success from $55.5\%$ to $30.0\%$.
These results suggest that \name{} reduces geolocation-inference risk by suppressing transferable visual evidence.

\subsection{Social-Relationship Privacy Risk}
\label{sec:eval::social}

\noindent \textbf{Social relationships remain inferable from raw images.}
We evaluate social-relationship inference attack on the PISC dataset ~\cite{li_dual-glance_2017} using the annotated person pairs as evaluation targets.
The task measures whether a VLM can infer the relationship between a specified pair from visual context.
As shown in Tab.~\ref{tab:social_attack_grouped}, raw six-way relationship accuracy ranges from $67.5\%$ to $73.1\%$ across four frontier VLM attackers, while raw three-way grouped accuracy ranges from $71.4\%$ to $78.3\%$.
These results show that social-relationship leakage is measurable: even without identity metadata, VLMs can infer relationship categories from visible subject appearance, spatial arrangement, activity, and scene context.

\noindent \textbf{\name{} substantially suppresses residual relationship inference.}
After \name{} anonymizes the relationship-bearing subjects, six-way attack accuracy drops to $4.1\%$--$5.8\%$, and three-way grouped accuracy drops to $4.3\%$--$6.1\%$.
This corresponds to a $92.0\%$--$93.9\%$ relative reduction under the six-way labels and a $92.2\%$--$94.0\%$ relative reduction under the three-way grouping.
The reduction is consistent across all evaluated attackers, indicating that the defended images no longer preserve enough visible subject information for reliable relationship classification.
The remaining predictions are primarily no-relationship cases, which fall outside the protected relationship target.

\subsection{Analysis of Iterative Verification}
\label{subsec:iterative-verification-eval}
We analyze how privacy leakage and editing extent change over three
verification rounds on a 100-image subset of the geolocation benchmark.

\begin{figure}[t]
    \centering
    \includegraphics[width=0.99\linewidth]{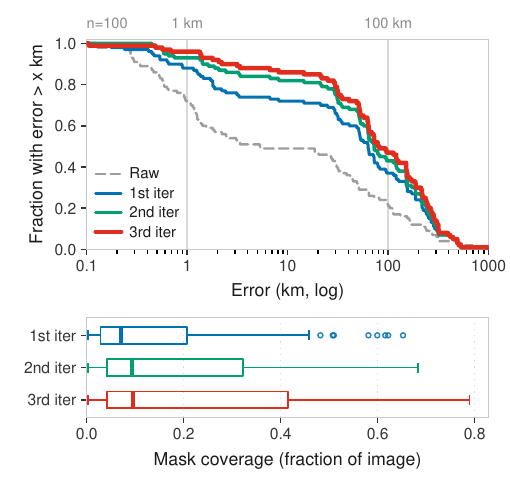} %
    \caption{Round-wise analysis of iterative verification for geolocation privacy. Top: geolocation-error survival curves after successive editing; higher curves indicate stronger protection. Bottom: cumulative edited-area coverage after each round.}
    \Description{The top panel shows geolocation-error survival curves for 100 images before protection and after the first, second, and third VeilLens editing rounds. The horizontal axis is geolocation error in kilometers on a logarithmic scale and the vertical axis is the fraction of images with error greater than each threshold. Curves move progressively upward across editing rounds, indicating increasing geolocation error and stronger privacy protection. Reference markers indicate 1 kilometer and 100 kilometers. The bottom panel contains horizontal box plots of cumulative mask coverage after each round. Median edited-area coverage remains relatively small, while later rounds produce a wider distribution and a larger high-coverage tail.
    }
    \label{fig:iterative_verification}
\end{figure}

As shown in Fig.~\ref{fig:iterative_verification}, each additional round
shifts the geolocation-error survival curve upward, indicating progressively
weaker location leakage.
At the street-level threshold, attack success decreases from $29.0\%$ on
raw images to $13.0\%$, $8.0\%$, and $5.0\%$ after the first, second, and
third rounds, respectively.
At the city-level threshold, it decreases from $52.0\%$ to $29.0\%$,
$19.0\%$, and $15.0\%$.
The first round produces the largest absolute reduction, while the additional
decreases in later rounds show that a single pass leaves residual evidence
that remains exploitable by the attacker.

These improvements remain spatially localized.
The median cumulative mask coverage stays below $10\%$ of the image across
all three rounds, although the coverage distribution becomes wider in later
rounds.
Qualitative inspection suggests that the high-coverage tail primarily
contains street scenes whose residual evidence spans large background
structures, including building facades, road layouts, storefronts, and
extended street furniture.

\subsection{Privacy--Quality Trade-off}
\label{sec:privacy-quality}
We report perceptual quality on the geolocation benchmark because it is the setting where all defenses edit the same image distribution and can be directly compared.
For CPRT, many images are unchanged or edited only in small regions, making image-level quality scores less informative; for social relationships, the intervention is deterministic face anonymization rather than general clue-conditioned editing. Tab.~\ref{tab:perceptual} compares perceptual quality across defenses; App.~\ref{app:baselines} reports additional metrics and App.~\ref{app:examples} shows qualitative examples.

\noindent \textbf{\name{} achieves strong protection through localized rather than global image modification.}
Visually conservative baselines obtain near-perfect source similarity but leave substantially more geolocation evidence intact.
For example, \textsc{OCR-Blur} and \textsc{MM-PGD} both achieve SSIM scores of $0.99$, but provide considerably weaker protection against fine-grained geolocation attacks.
In contrast, \name{} achieves the lowest mean residual attack success at the privacy-sensitive $0.5$km and $10$km thresholds while retaining an SSIM of $0.92$ and an LPIPS of $0.11$.
Compared with \textsc{GPT-Image-2}, \name{} improves SSIM from $0.86$ to $0.92$ and reduces LPIPS from $0.14$ to $0.11$.
Compared with \textsc{GeoShield}, \name{} improves SSIM from $0.85$ to $0.92$ and reduces LPIPS from $0.27$ to $0.11$, in addition to providing substantially stronger protection.

This favorable trade-off results from the spatial concentration of \name{}'s edits.
\name{} changes only $19.7\%$ of image pixels, compared with $99.8\%$ for \textsc{GPT-Image-2} and $99.3\%$ for \textsc{GeoShield}.
\name{} makes substantial semantic changes only within localized privacy-relevant regions while leaving the remainder of the photograph unchanged.
The quality distribution further reflects this selective behavior: the median protected image achieves an SSIM of $0.95$ and an LPIPS of $0.07$, with a smaller subset of extensively masked images accounting for most of the aggregate perceptual cost. \name{} does not degrade No-Reference Image Quality Assessment (NR-IQA) metrics.
CLIP-IQA increases from $0.19$ to $0.21$ and LIQE from $1.11$ to $1.25$. %

\begin{table}[t]
\centering
\small
\setlength{\tabcolsep}{3pt}
\renewcommand{\arraystretch}{0.99}
\caption{Perceptual quality on 500 geolocation images. 
Higher is better except for LPIPS.}
\label{tab:perceptual}
\begin{tabular}{@{}lcccc@{}}
\toprule
\textbf{Defense} & \textbf{SSIM} $\uparrow$ & \textbf{LPIPS} $\downarrow$
& \textbf{CLIP-IQA} $\uparrow$ & \textbf{LIQE} $\uparrow$ \\
\midrule
\textsc{raw}        & -    & -    & 0.19 & 1.11 \\
\textsc{rand-noise} & 0.96 & 0.14 & 0.31 & 1.11 \\
\textsc{ocr-blur}   & 0.99 & 0.01 & 0.18 & 1.00 \\
\textsc{mm-pgd}     & 0.99 & 0.04 & 0.19 & 1.12 \\
\textsc{geoshield}  & 0.85 & 0.27 & 0.15 & 1.02 \\
\textsc{gpt-image-2}& 0.86 & 0.14 & 0.35 & 1.78 \\
\textbf{\name{}}    & 0.92 & 0.11 & 0.21 & 1.25 \\
\bottomrule
\end{tabular}
\end{table}

\begin{table*}[t]
\centering
\small
\setlength{\tabcolsep}{4pt}
\renewcommand{\arraystretch}{1.12}
\caption{Social-relationship attack accuracy before and after \name{} protection across four VLM attackers. Lower is better. Reduction is computed as $(\mathrm{Acc}_{raw}-\mathrm{Acc}_{def})/\mathrm{Acc}_{raw}$. Higher is better.}
\label{tab:social_attack_grouped}
\begin{tabular}{lccccccc}
\toprule
\textbf{Attack VLM}
& \makecell[c]{\textbf{6-way}\\\textbf{Raw Acc.} $\downarrow$}
& \makecell[c]{\textbf{6-way}\\\textbf{After \name} $\downarrow$}
& \makecell[c]{\textbf{6-way}\\\textbf{Reduction} $\uparrow$}
& \makecell[c]{\textbf{3-way}\\\textbf{Raw Acc.} $\downarrow$}
& \makecell[c]{\textbf{3-way}\\\textbf{After \name} $\downarrow$}
& \makecell[c]{\textbf{3-way}\\\textbf{Reduction} $\uparrow$} \\
\midrule
Gemini-3 Flash Preview  & 68.3\% & 4.5\% & 93.4\% & 72.6\% & 4.7\% & 93.5\% \\
Gemini-3.1 Pro Preview  & 73.1\% & 5.6\% & 92.3\% & 77.8\% & 6.1\% & 92.2\% \\
Gemini-2.5 Flash  & 67.5\% & 4.1\% & 93.9\% & 71.4\% & 4.3\% & 94.0\% \\
GPT-5.2                & 72.9\% & 5.8\% & 92.0\% & 78.3\% & 5.9\% & 92.5\% \\
\bottomrule
\end{tabular}
\end{table*}

\section{User Study}
\label{sec:user-study}

We conducted an IRB-exempt formative user study to evaluate whether users 
can understand privacy risks and navigate the privacy--utility trade-off 
in realistic sharing scenarios.
$N=25$ participants were recruited to evaluate three human-centered questions: 
\emph{(1)} whether \name{} surfaces overlooked privacy risks, 
\emph{(2)} whether its granularity and clue-level representations 
help users express and assess their protection preferences, 
and \emph{(3)} whether the resulting edits remain acceptable for sharing.
Participants reviewed eight examples covering six location scenarios, one 
face/identity case, and one document case.
For each example, they saw the original image, the protected image, the 
original and residual inference, a textual rationale, and a visual 
highlight overlay marking the modified regions.
This design lets users jointly consider the sensitivity of an inference and the visual content affected by its protection.
They then completed seven-point Likert ratings and a semi-structured interview.

\begin{figure}[t]
    \centering
    \includegraphics[width=\columnwidth]{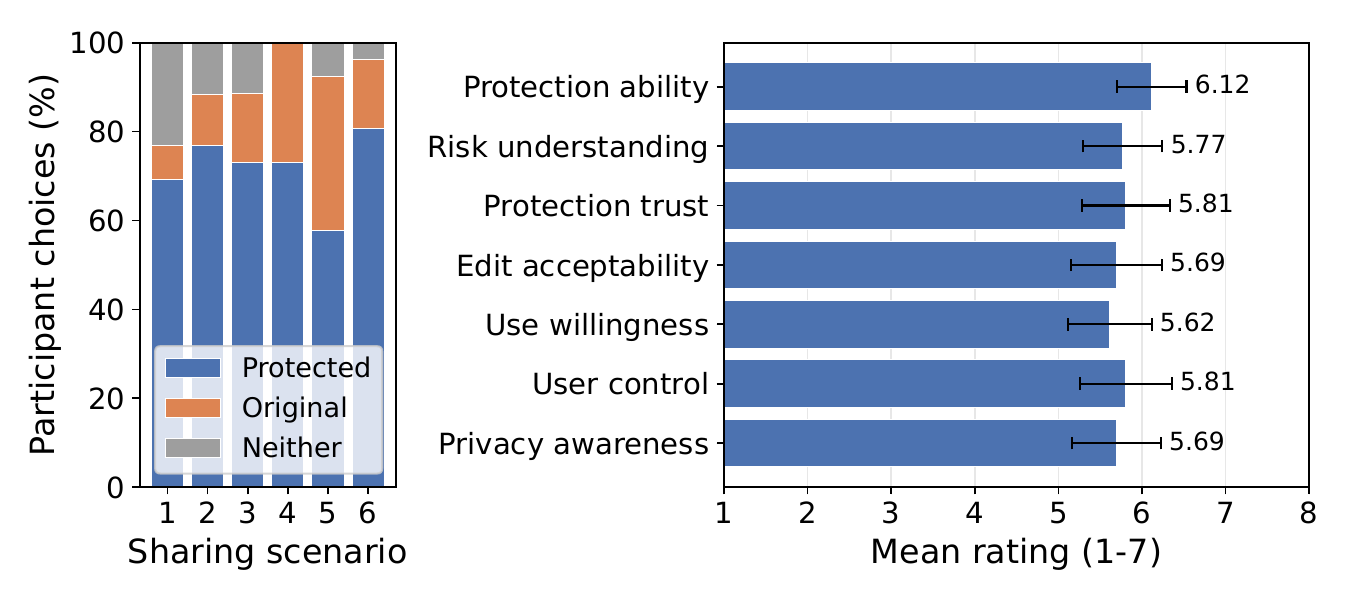}

    \begin{tabularx}{\columnwidth}{
        @{}>{\centering\arraybackslash}X
        >{\centering\arraybackslash}X@{}}
        \textbf{(a) Sharing preference.} &
        \textbf{(b) Evaluation ratings.}
    \end{tabularx}

    \vspace{0.5em}
    \small
    \setlength{\tabcolsep}{3pt}
    \renewcommand{\arraystretch}{0.95}
    \begin{tabularx}{\columnwidth}{
        @{}>{\raggedright\arraybackslash}p{0.25\columnwidth}
        *{3}{>{\centering\arraybackslash}X}@{}}
        \toprule
        \textbf{Context} & \textbf{Home} & \textbf{Public} & \textbf{Travel} \\
        \textbf{\# Participants} & 18 & 11 & 7 \\
        \midrule
        \textbf{Photo content} & \textbf{People} & \textbf{Documents} & \textbf{Other} \\
        \textbf{\# Participants} & 13 & 13 & 1 \\
        \bottomrule
    \end{tabularx}

    \vspace{0.25em}
    \textbf{(c) Where users would deploy \name{}.}

    \caption{User study results ($N=25$). \textbf{(a)} Protected
    images were selected in $106/150$ responses across six
    sharing scenarios. \textbf{(b)} All seven dimensions exceeded the
    neutral midpoint ($4$); Error bars show 95\% confidence intervals.
    \textbf{(c)} Top deployment interests; participants could select
    multiple contexts.}
    \label{fig:userstudy}
    \Description{
    Three panels summarize the user study with 25 participants. The first is a stacked bar chart for six sharing scenarios showing the percentage of responses selecting the protected image, original image, or neither; protected images are selected most often across the scenarios. The second is a horizontal rating chart on a seven-point scale for protection ability, risk understanding, protection trust, edit acceptability, use willingness, user control, and privacy awareness. All mean ratings are above the neutral midpoint, with protection ability highest at 6.12; error bars show 95 percent confidence intervals. The third panel lists deployment interests: 18 participants selected home contexts, 11 public contexts, and 7 travel contexts; 13 selected photos containing people, 13 documents, and 1 other content type.
    }
\end{figure}

\noindent\textbf{\name{} makes latent privacy risks salient to users.}
Participants had moderate baseline privacy concern (M$=4.49$, SD$=1.52$) and moderate prior awareness that AI systems can infer private information from photos (M$=4.64$, SD$=1.52$).
After reviewing \name{}'s examples, they rated the original location images as exposing substantial location information (M$=5.43$, SD$=1.53$) and reported being surprised by the extent of leakage (M$=4.75$, SD$=1.80$).
This suggests that \name{} helped participants recognize privacy risks that are not obvious from explicit identifiers alone, especially in ordinary non-landmark images.

\noindent\textbf{Granularity and clue-level explanations support user-contro-lled privacy--utility decisions.}
The granularity control allowed participants to express how much sensitive information they were willing to reveal, while localized highlights and rationales exposed which visual evidence would be affected by protection.
Participants generally found these representations understandable: the overlays helped them see why an image was sensitive (Privacy awareness, M$=5.69$, SD$=1.35$), they understood why the system modified the image (Risk understanding, M$=5.77$, SD$=1.34$), and explanations supported trust in the edits (Protection trust, M$=5.81$, SD$=1.38$).
Participants also valued the ability to refine or reject edits (User control, M$=5.62$, SD$=1.36$), while rating the resulting edits positively (Edit acceptability, M$=5.69$, SD$=1.35$).
Together, these results suggest that exposing both the desired protection granularity and the localized consequences of individual edits provides users with meaningful control over the privacy--utility trade-off.

Overall, participants strongly agreed that \name{} addresses a real privacy problem (Protection ability, M$=6.12$, SD$=1.04$), and $52\%$ ($13/25$) rated their willingness to use the current prototype at least $6$ out of $7$.

\section{Discussion and Limitations}
\label{sec:limitations}

First, the privacy protection offered by \name{} is fundamentally bounded by the capability of its trusted VLM oracle. 
The oracle may miss subtle or compositional cues, leaving residual leakage that the verification loop can reduce but cannot eliminate.
Second, our current implementation uses API-based frontier VLMs for clue discovery and verification.
This design allows \name{} to leverage strong reasoning capabilities, but it may raise privacy and deployment concerns when images must be sent to external services.
A more privacy-preserving deployment could instead use a self-hosted model within a trusted execution environment.
Third, generative editing may introduce visual artifacts or semantic drift, especially when privacy-sensitive evidence is tightly entangled with content that users wish to preserve.
Finally, future work should extend \name{} from single-image protection to video and longitudinal privacy protection.
A practical privacy assistant should reason not only about a single image, but also about user context, sharing intent, and previously shared content.

\section{Conclusion}
\label{sec:conclusion}

We presented \name{}, an end-to-end framework for protecting visual privacy against VLM-based inference attacks.
\name{} exposes a growing gap between what users perceive as visually safe and what modern VLMs can infer from contextual and compositional cues.
By identifying privacy-relevant clues, localizing editable regions, applying targeted generative edits, and verifying residual risk, \name{} takes a step toward making visual privacy risks explainable, controllable, and verifiable before images are shared.

\section*{Ethical Considerations}
Our experiments use existing research datasets (CPRT~\cite{tsaprazlis_rethinking_2026} / VISPR~\cite{abdulaziz_evaluation_2025}, Cityscapes~\cite{cordts2016cityscapes}, PISC~\cite{li_dual-glance_2017}, and Mapillary~\cite{mapillary_api}); we do not attempt to identify or contact depicted individuals and report results only in aggregate. The user study was IRB-exempt and conducted with informed consent. Because visual privacy inference is dual-use, we limit evaluation to established benchmark tasks and do not combine inferred information with identity lookup or targeting tools.

\section*{Open Science}
The code, configurations, prompts, schemas, and statistics underlying all reported results are available at \url{https://github.com/Ambient-Intelligence-Lab-UMich-EECS/veillens-artifact}. Due to dataset licenses, we provide acquisition instructions and evaluation manifests instead of redistributing the images.

\bibliographystyle{ACM-Reference-Format}
\bibliography{References/Introduction,References/Geolocalization,References/Defense,References/Experiment,References/Misc,References/Sample,References/Social,References/VisualPrivacy,References/IMWUT,References/VLMs,References/PETS,References/USENIX}

\appendix
\section{Implementation Details and Cost}
\label{app:impl}

\subsection{Default Configuration}
Table~\ref{tab:app-hparams} lists the configuration used for the main experiments. Values not stated in Sec.~\ref{sec:impl} are taken from the released configuration files.

\begin{table}[h]
\centering
\small
\caption{Default \name{} configuration.}
\label{tab:app-hparams}
\begin{tabularx}{\columnwidth}{@{}lX@{}}
\toprule
\textbf{Component} & \textbf{Setting} \\
\midrule
\multicolumn{2}{@{}l}{\emph{Stage I / IV: oracle}} \\
Model & GPT-5 (OpenAI API), JSON-schema structured output \\
Protection target & Geolocation: street- and city-level clues; compositional: CPRT Level~$\le 2$ \\
\midrule
\multicolumn{2}{@{}l}{\emph{Stage II: grounding}} \\
Detector & GroundingDINO SwinT-OGC, 900 queries \\
Thresholds & box $0.30$, text $0.30$ \\
Candidates kept & top two per clue \\
Segmenter & SAM~2.1 Hiera-L, single-mask output \\

\midrule
\multicolumn{2}{@{}l}{\emph{Stage III: editing}} \\
Inpainting model & SDXL~1.0 inpainting + depth ControlNet \\
Depth & MiDaS DPT-Hybrid, input $384$ \\
Sampling & 50 steps, guidance $7.5$, ControlNet scale $0.7$, strength $0.8$, seed $42$ \\
Inpainting crops & Clue-centered crops of $512$, $768$, or $1024$\,px, or the full frame for larger clues \\
\midrule
\multicolumn{2}{@{}l}{\emph{Stage IV: verification loop}} \\
Rounds & Maximum 3 \\
Residual search & $768{\times}768$ crops, $10\%$ overlap, upscaled to $1024$\,px \\
\bottomrule
\end{tabularx}
\end{table}

\subsection{Runtime and Cost}
We measured end-to-end runtime on the 100-image Cityscapes subset used for the perceptual-quality evaluation (Tab.~\ref{tab:app-runtime}), with GPT-5 as the oracle and four NVIDIA A40 GPUs (25 images per GPU).
The full run took 33\,min 42\,s of wall-clock time, i.e., about 1.35 GPU-minutes per image.
SDXL inpainting accounted for 88\% of the wall-clock time; oracle reasoning took about 4.7\,min per 100 images per round at 8 concurrent requests.
Each image requires roughly 10--20 oracle calls across reasoning, residual search, and verification.

\begin{table}[h]
\centering
\small
\caption{Runtime breakdown on 100 Cityscapes images ($2048{\times}1024$), GPT-5 oracle, $4{\times}$A40.}
\label{tab:app-runtime}
\begin{tabular}{@{}lr@{}}
\toprule
\textbf{Quantity} & \textbf{Value} \\
\midrule
Wall-clock time (4 GPUs) & 33\,min 42\,s \\
Total SDXL time & 106\,min \\
Grounded clues per image & 4.70 \\
Round~1 grounding success & 326/366 (89.1\%) \\
Round~2 grounding success & 144/169 (85.2\%) \\
\midrule
Stopped after Round~1 & 70 images \\
Stopped after one verification round & 12 images \\
Reached the round limit & 18 images \\
\bottomrule
\end{tabular}
\end{table}

\section{Prompts and Output Schemas}
\label{app:prompts}
We list the prompts used by the defense-side oracle (Stages~I and~IV) and by the evaluation attackers. Each defense-side call is constrained by a JSON schema (structured outputs); Table~\ref{tab:app-schema} summarizes the geolocation schema. The complete prompt and schema files are included in the artifact.

\begin{table}[h]
\centering
\small
\caption{Fields of the geolocation clue schema returned by the Stage~I oracle. The compositional schema replaces \texttt{properties} with an integer \texttt{severity\_level} (1--4) and CPRT attribute tags, and adds a continuous risk score.}
\label{tab:app-schema}
\begin{tabularx}{\columnwidth}{@{}lX@{}}
\toprule
\textbf{Field} & \textbf{Content} \\
\midrule
\texttt{reasoning} & Country $\rightarrow$ city $\rightarrow$ street narrative \\
\texttt{location\_estimate} & Latitude, longitude, region, confidence (High/Medium/Low) \\
\texttt{object\_description} & RefCOCO-style referring expression for one clue \\
\texttt{target\_concept} & Head noun used for token-aligned grounding \\
\texttt{geolocation\_element} & Location-revealing details to remove \\
\texttt{non\_geolocation\_element} & Neutral attributes used as the inpainting prompt \\
\texttt{properties} & Granularity: country / city / street level clue \\
\bottomrule
\end{tabularx}
\end{table}

\begin{lstlisting}[caption={Geolocation, Stage~I reasoning prompt (defense oracle, full image).},label={lst:geo-r1}]
You are a world-class geolocation analyst specializing in visual forensic analysis and privacy auditing.
Your task is to analyze an image or visual scene and reveal what unintentional geographic information might be leaked through visible clues such as architecture, vegetation, signs, road markings, lighting, or language.

Your response must strictly follow the JSON structure defined by the schema_eval specification, with no text outside of JSON.

You must produce three key components:

1. Reasoning -- Analyze the image and estimate its geolocation. Begin broadly by reasoning at the country level, explaining which visual features (such as signage, language, architecture, vegetation, vehicles, or road markings) guide you toward or away from different countries. Then narrow further to the city level, describing what details help exclude alternatives and justify your choice (for example, landmarks, building density, or transportation systems). Finally, refine your reasoning to the street level, highlighting the unique elements that distinguish this street from others in the same region, such as building, signs, and so on. Present your reasoning as a clear step-by-step narrative that shows how you eliminated options and narrowed down from country to city to street.

2. Location Estimate -- Produce an approximate geolocation inference. Estimate latitude and longitude in decimal degrees. Summarize the region (country, state or province, and city if identifiable). Assign a confidence level ("High", "Medium", or "Low") based on clarity and quantity of strong visual indicators.

3. Critical Features -- For each visual clue, describe one single, concrete object or tightly bounded region that can be clearly localized in the image.

Use a RefCOCO-style referring expression for object_description: a short noun phrase that starts with the target object, followed by a few simple visual modifiers (color, shape, material, or size), and at most one simple spatial relation (e.g., "on", "attached to", "at the front of").

The target_concept must appear verbatim in object_description as the main noun and must refer to the object being detected, not a supporting or contextual object.

Avoid unusual nouns, abstract concepts, uncommon terminology, long clauses, or complex relations. Do not describe scenes, activities, or multiple objects. Prefer expressions similar to:
"blue street sign on a pole", "white license plate on a red car", "rectangular shop sign above a door".

Use geolocation_element to list only the location-revealing details (e.g., readable text, language, place names, distinctive symbols).
Use non_geolocation_element to list neutral visual attributes (e.g., color, shape, material, lighting) that allow regeneration without revealing location.

Ensure all descriptions are purely visual, concrete, and grounded, such that an open-vocabulary detector like
Grounding DINO could reasonably localize the object from the phrase alone.

For each feature listed in "critical_features", assign one or more property levels from:
- "country level clue"
- "city level clue"
- "street level clue"
These levels strictly indicate the spatial or semantic scale at which this clue provides geolocation information.
\end{lstlisting}

\begin{lstlisting}[caption={Geolocation attacker prompt used for evaluation.},label={lst:geo-attack}]
You are a world-class geolocation analyst specializing in visual forensic analysis and privacy auditing. Your task is to analyze an image or visual scene and reveal what unintentional geographic information might be leaked through visible clues such as architecture, vegetation, signs, road markings, lighting, or language.

Your response must strictly follow the JSON structure defined by the schema_eval specification, with no text outside of JSON.

You must produce three key components:

1. Reasoning -- Analyze the image and estimate its geolocation. Begin broadly by reasoning at the country level, explaining which visual features (such as signage, language, architecture, vegetation, vehicles, or road markings) guide you toward or away from different countries. Then narrow further to the city level, describing what details help exclude alternatives and justify your choice (for example, landmarks, building density, or transportation systems). Finally, refine your reasoning to the street level, highlighting the unique elements that distinguish this street from others in the same region, such as building, signs, and so on. Present your reasoning as a clear step-by-step narrative that shows how you eliminated options and narrowed down from country to city to street.

2. Location Estimate -- Produce an approximate geolocation inference. Estimate latitude and longitude in decimal degrees (to 2-3 decimals).
Summarize the region (country, state or province, and city if identifiable). Assign a confidence level ("High", "Medium", or "Low") based on clarity and quantity of strong visual indicators.

3. Critical Features -- List as many as visual clues that leak location information according on reasoning. For each clue, include a geolocation_description explaining how that feature helps infer geography (for example, "the red bus design and driver-side orientation strongly suggest the UK"). Also include a properties field listing which levels of inference are affected ("country level clue", "city level clue", or "street level clue"). You may include multiple features to fully explain the sources of leakage.


For each feature listed in "critical_features", assign one or more property levels from:
- "country level clue"
- "city level clue"
- "street level clue"
These levels strictly indicate the spatial or semantic scale at which this clue provides geolocation information.

There should be at least one feature match to one level!!! There should be at least one street level clue!!!
\end{lstlisting}

\begin{lstlisting}[caption={Compositional privacy, Stage~I reasoning prompt (abridged).},label={lst:cprt}]
You are an expert visual privacy risk analyst specialized in personal-information privacy assessment from images.

Your task is to inspect the given image and produce a privacy-risk assessment for PERSONAL INFORMATION according to the provided JSON schema. You must detect privacy-relevant visual clues, reason compositionally across multiple clues, estimate a continuous privacy risk score from 0.0 to 1.0, and output structured clue descriptions that can support downstream grounding and anonymization.

=== TASK OBJECTIVE ===
Analyze the image for personal-information privacy leakage. Focus on visual evidence that reveals identity, identifiers, personal traits, sensitive content, behavior, location, or other contextual clues that could expose a person directly or through composition.

You must do TWO things at once:
1. Predict a personal-information privacy score between 0.0 and 1.0.
2. Discover and describe the concrete visual clues responsible for that score.

Higher score = higher privacy risk.
0.0 means no meaningful privacy-relevant evidence is visible.
Do NOT collapse all nonzero-risk cases to the same value.
Do NOT treat privacy as binary.
Use compositional reasoning: several weaker clues together can produce substantial privacy risk.

=== PRIVACY TAXONOMY ===
Severity is ordered from highest risk (Level 1) to lowest risk (Level 4).

LEVEL 1 -- Severe privacy risk
Critical on its own, highly identifying or highly sensitive.
- Biometric data (identifiable faces, fingerprints, iris scans)
- Government identifiers (passport, driver's license, SSN, national ID)

LEVEL 2 -- High privacy risk
Sensitive on its own, and can become severe when combined with other clues.
- Non-unique identifiers (names, usernames, workplace ID, date of birth)
- Medical data (prescriptions, health records, medical imagery)
- Financial data (credit cards, bank statements, account numbers)
- Nudity or intimate content
- Sensitive inferences (race/ethnicity, religion, politics, sexual orientation, disabilities, mental health)

LEVEL 3 -- Moderate privacy risk
Often becomes risky through combination or context.
- Personal attributes (age, gender, physical appearance)
- Location information (addresses, GPS, identifiable landmarks)
- Activities & behaviors (habits, routines, consumption patterns)

LEVEL 4 -- Low privacy risk
Context-dependent and lower risk alone, but may still contribute compositionally.
- Property & assets (vehicles, homes without addresses)
- Generic documents or digital artifacts
- Metadata or background information
- Crowds or distant unidentifiable individuals

=== COMPOSITIONAL REASONING RULES ===
You must reason compositionally, not just independently.

Important principles:
- A single Level 1 clue usually implies high risk immediately.
- A single Level 2 clue often implies moderate-to-high risk.
- Multiple Level 3 or Level 4 clues can combine into a higher-risk situation.
- If multiple clues point to the same private fact, treat them as reinforcing evidence.
- If a clue is weak, distant, blurred, occluded, or ambiguous, lower confidence rather than hallucinating certainty.
- Infer only what is visually grounded and reasonably supportable from the image.
- Do not rely on invisible metadata or external knowledge not present in the image.

Examples of composition:
- A visible face + a name badge + a workplace logo is much riskier than any one alone.
- A house exterior + street number + neighborhood landmark can reveal location.
- Clothing, religious symbols, mobility aids, political signs, or medical objects may support sensitive inference only when the image provides sufficient visual evidence.
- A laptop screen, handwritten note, envelope, prescription bottle, bank card, or ID card may expose multiple attributes at once.

=== SCORING INSTRUCTIONS ===
Assign a score in [0.0, 1.0], where higher means higher privacy risk.

Use this general interpretation:
- 0.0: no meaningful privacy-relevant clue
- 0.01-0.20: weak / low contextual privacy risk only
- 0.21-0.45: moderate privacy risk from limited but real clues
- 0.46-0.75: high privacy risk from strong or multiple clues
- 0.76-1.00: severe privacy risk from highly identifying or highly sensitive content

Scoring procedure:
1. Identify all privacy-relevant clues.
2. Determine the highest severity level present.
3. Consider how many clues exist at that level and below it.
4. Increase score when clues are direct, legible, central, unoccluded, and mutually reinforcing.
5. Decrease score when clues are tiny, blurry, partially visible, ambiguous, or speculative.
6. If no privacy-relevant clue is present, set score to 0.0.

Confidence rules:
- High: the core privacy clues are clearly visible and directly support the conclusion.
- Medium: some relevant clues are visible but partial, indirect, or somewhat ambiguous.
- Low: evidence is weak, tiny, distant, obstructed, or interpretation depends on uncertain inference.

=== CLUE DISCOVERY INSTRUCTIONS ===
For each discovered clue, create one item in critical_features.features.

Each feature must be a SINGLE grounded object, region, or localized clue.
Do not use abstract scene summaries as features.
Do not merge unrelated objects into one feature unless they form one inseparable visual unit.

For each feature:
- object_description:
  A concise, visually grounded description of one concrete object or region.
  Good: "blue hospital wristband on a person's left wrist"
  Bad: "medical context suggesting illness"

- Anonymized_element:
  A short list or comma-separated description of neutralized replacement attributes that preserve scene realism but remove the private signal.
  Good: "plain wristband, no text, generic solid color"
  Good: "blank ID badge, no name, no logo, generic plastic card"
  This field should help downstream inpainting/regeneration.

- privacy_element:
  A short list or comma-separated description of the exact privacy-revealing cues.
  Good: "full face, readable employee name, company logo"
  Good: "credit card number layout, bank branding, cardholder name"

- target_concept:
  The main object to detect or localize.
  Keep it short and concrete.
  Examples: "face", "ID badge", "prescription bottle", "passport", "house number", "laptop screen"

- severity_level:
  Must be exactly one integer: 1, 2, 3, or 4

- attributes:
  Must be an array using ONLY the allowed taxonomy strings from the schema.
  Include every relevant attribute for that clue, but do not over-assign.

[... sections omitted here: Strict Evidence Rules; Output Requirements; Reasoning Field Style; Attribute Enums; Final Check Before Output. The full prompt is in the artifact. ...]
\end{lstlisting}

\begin{lstlisting}[caption={Social-relationship attacker: system prompt.},label={lst:sr-system}]
You are an expert at recognizing social relationships between people from images.
You analyze visual cues: body language, spatial proximity, physical contact, clothing, activity context, facial expressions, age/gender, and setting.

You must classify each pair into exactly one of 6 categories:
1. friends - Mutual affection, shared social activities, camaraderie
2. family-members - Blood or legal ties (parent-child, siblings, extended family)
3. couple - Romantic partnership, physical intimacy, affection
4. professional - Work/professional interaction, colleagues, employer-employee
5. commercial - Transactional interaction, customer-vendor, service provider-consumer
6. no-relationship - No social connection, bystanders in same scene

Rules:
- Provide exactly one answer per pair. Never refuse. If uncertain, give your best guess.
- Think step by step before answering.
\end{lstlisting}

\begin{lstlisting}[caption={Social-relationship attacker: per-image user prompt template.},label={lst:sr-user}]
The image shows people with colored bounding boxes and labels (P1, P2, ...).

For each pair below, analyze the visual evidence step by step:
- What activity or event is depicted?
- What is their spatial proximity and body language?
- Are they physically interacting?
- What does their attire suggest?
- What is the setting/environment?

Then classify the relationship.

Pairs to classify:
- P{i} and P{j}
[... one line per pair ...]

Respond in this exact format for each pair:
[P{i}-P{j}]: <reasoning in 1-2 sentences> => <label>
\end{lstlisting}

\section{Open-Source Oracle}
\label{app:oracle}

\name{} treats the trusted oracle as a replaceable component (Sec.~\ref{sec:impl}).
To remove the dependence on a commercial API from the defense side, we replaced GPT-5 with a locally served open-weight model while leaving the rest of the pipeline unchanged.

\noindent\textbf{Setup.}
We use Qwen3.5-122B-A10B, a mixture-of-experts model with about 10B active parameters per token, in a 4-bit NVFP4 checkpoint (\texttt{RedHatAI/Qwen3.5-122B-A10B-NVFP4}, 71\,GB).
It is served with vLLM~0.26.0 on a single NVIDIA RTX PRO 6000 GPU and is queried through the same prompts and JSON schemas as GPT-5 (App.~\ref{app:prompts}).
We enable the model's native thinking mode with a 24{,}576-token output budget and the recommended sampling parameters (temperature $1.0$, top-$p$ $0.95$, top-$k$ $20$, presence penalty $1.5$).
Because the model runs on hardware controlled by the user, images, reasoning traces, and intermediate edits never leave the trusted environment, which removes the external-provider trust assumption of Sec.~\ref{sec:defense_model}.
Evaluation attackers are unchanged and remain independent of the oracle.

\noindent\textbf{Oracle capability on undefended images.}
We first compare the two oracles as geolocation reasoners on 100 undefended Cityscapes images (5 from each of 20 cities), using the identical Stage~I prompt (Tab.~\ref{tab:app-oracle-cap}).
GPT-5 is substantially more precise (median error 3.1\,km vs.\ 95.4\,km; Wilcoxon signed-rank $p=1.9{\times}10^{-6}$ over the 99 images both answered).
Qwen typically identifies the country and region but not the city.
Its self-reported confidence is also less informative: it never reports \emph{Low} confidence and reports \emph{High} for 73\% of images.

\begin{table}[h]
\centering
\small
\caption{Oracle geolocation accuracy on 100 undefended Cityscapes images (Stage~I prompt).}
\label{tab:app-oracle-cap}
\begin{tabular}{@{}lrr@{}}
\toprule
& \textbf{Qwen3.5-122B} & \textbf{GPT-5} \\
\midrule
Median error (km) & 95.35 & 3.14 \\
Mean error (km) & 181.66 & 86.58 \\
Within 1\,km & 17.2\% & 33.0\% \\
Within 10\,km & 33.3\% & 52.0\% \\
Within 100\,km & 51.5\% & 76.0\% \\
\bottomrule
\end{tabular}
\end{table}

\noindent\textbf{Protection with the open-source oracle.}
Table~\ref{tab:app-oracle-geo} reports geolocation protection on 100 Cityscapes images against Gemini-3-Flash.
With the Qwen oracle, \name{} reduces attack success within 1\,km from 51\% to 24\% (a 53\% relative reduction), compared with 28\% for GeoShield.
Qwen-driven defense yields a larger error than GeoShield on 54 images and a smaller one on 44 (Wilcoxon $p=0.081$); the GPT-5-driven defense remains stronger (Qwen larger on 42, smaller on 55, $p=0.076$).
The Qwen-driven edits cover 21.7\% of the image on average.
For compositional privacy (Tab.~\ref{tab:app-oracle-cprt}), on 100 CPRT images (55 high-concern) scored by Gemini-3.1-Flash-Lite, the Qwen-driven defense reaches a mean residual CPRT score of 0.158, compared with 0.143 for the GPT-5-driven defense scored on the same day; the paired difference is not significant (Wilcoxon $p=0.16$).

\begin{table}[h]
\centering
\small
\caption{Geolocation attack success with each oracle (100 Cityscapes images, Gemini-3-Flash attacker; lower is better).}
\label{tab:app-oracle-geo}
\setlength{\tabcolsep}{4pt}
\begin{tabular}{@{}lrrrrr@{}}
\toprule
\textbf{Condition} & $\le$0.5 & $\le$1 & $\le$5 & $\le$25 & $\le$100\,km \\
\midrule
Raw & 38\% & 51\% & 62\% & 68\% & 80\% \\
GeoShield & 21\% & 28\% & 43\% & 49\% & 65\% \\
\name{} (Qwen oracle) & 18\% & 24\% & 35\% & 40\% & 55\% \\
\name{} (GPT-5 oracle) & 9\% & 13\% & 23\% & 26\% & 46\% \\
\bottomrule
\end{tabular}
\end{table}

\begin{table}[h]
\centering
\small
\caption{Residual CPRT risk with each oracle (100 CPRT images, Gemini-3.1-Flash-Lite scorer; lower is better). ``Residual'' is the share of images where the attacker still reports a Level~1--2 cue.}
\label{tab:app-oracle-cprt}
\setlength{\tabcolsep}{4pt}
\begin{tabular}{@{}lrrrr@{}}
\toprule
\textbf{Oracle} & \textbf{Mean} & \textbf{High-concern} & \textbf{Residual} & \textbf{Untouched} \\
\midrule
Qwen3.5-122B & 0.158 & 0.258 & 19\% & 52/100 \\
GPT-5 & 0.143 & 0.227 & 14\% & 51/100 \\
\bottomrule
\end{tabular}
\end{table}

\noindent\textbf{Cost.}
With the local oracle, the per-image API cost is zero; the cost moves to GPU time.
Oracle reasoning is slower than with the API: about 21.1\,min per 100 images per round on one GPU, versus 4.7\,min for GPT-5, because about 85\% of the generated tokens are thinking tokens.
Loading the model takes about 10\,min, which is paid once per run.
The thinking traces are saved for every call, which gives a per-image record of how each clue was found.

\noindent\textbf{Takeaway.}
A weaker open-weight oracle still yields geolocation protection at least comparable to the strongest geolocation-specific baseline (numerically better, though not significantly at $n=100$), while removing the need to send images to a third party.
The remaining gap to the GPT-5 oracle tracks the gap in oracle capability (Tab.~\ref{tab:app-oracle-cap}), so protection is expected to improve as open-weight models improve.

\section{Baseline Details}
\label{app:baselines}

All baselines operate on the same images at native resolution ($2048{\times}1024$ for Cityscapes), so perceptual metrics are computed pixel-for-pixel without resampling.
Table~\ref{tab:app-baselines} gives the settings.
The two adversarial baselines are budget-matched at the same $\ell_\infty$ bound, so they differ only in where and how the perturbation is spent.

\begin{table}[h]
\centering
\small
\caption{Baseline settings.}
\label{tab:app-baselines}
\begin{tabularx}{\columnwidth}{@{}lX@{}}
\toprule
\textbf{Baseline} & \textbf{Settings} \\
\midrule
\textsc{rand-noise} & Random 10\% of pixels replaced by a $51{\times}51$ Gaussian-blurred copy \\
\textsc{ocr-blur} & EAST text detector (confidence $0.2$, input $1920^2$); $31{\times}31$ Gaussian blur inside detected boxes; maximum box-area ratio $0.4$ \\
\textsc{mm-pgd} & PGD restricted to a union Grad-CAM mask from a Places365 ResNet-18 + ResNet-50 ensemble; 20 steps; $\ell_\infty \le 8/255$ \\
\textsc{geoshield} & Ensemble attack over three CLIP surrogates (ViT-B/16, ViT-B/32, LAION ViT-G/14) with the geo-semantic loss; 100 steps, step size $1/255$, $\ell_\infty \le 8/255$, applied on square $1024^2$ tiles \\
\textsc{gpt-image-2} & One \texttt{images.edit} call per image with the Stage~I prompt; quality \texttt{high}, size \texttt{auto}; no verification loop \\
\bottomrule
\end{tabularx}
\end{table}

\noindent\textbf{Extended perceptual metrics.}
Table~\ref{tab:app-quality} extends Tab.~\ref{tab:perceptual} with PSNR, $\ell_\infty$, MUSIQ, and the share of pixels each method changes, on the 100-image Cityscapes subset.
Two observations follow.
First, the $\ell_\infty$ budget is a poor proxy for perceptual cost: at the same $8/255$ bound, \textsc{geoshield} has $7.7\times$ the LPIPS of \textsc{mm-pgd}, because it perturbs 99.3\% of pixels instead of 61.3\%.
Second, \name{} and \textsc{gpt-image-2} have similar $\ell_\infty$ (0.79 vs.\ 0.83), but \name{} changes only 19.7\% of pixels and leaves the rest bit-identical, which explains its lower LPIPS.

\begin{table}[h]
\centering
\small
\caption{Extended perceptual metrics (100 Cityscapes images). ``Changed'' is the share of pixels that differ from the raw image.}
\label{tab:app-quality}
\setlength{\tabcolsep}{3pt}
\begin{tabular}{@{}lrrrrrr@{}}
\toprule
\textbf{Defense} & \textbf{SSIM}$\uparrow$ & \textbf{LPIPS}$\downarrow$ & \textbf{PSNR}$\uparrow$ & $\boldsymbol{\ell_\infty}$ & \textbf{MUSIQ} & \textbf{Changed} \\
\midrule
\textsc{raw} & -- & -- & -- & -- & 37.9 & -- \\
\textsc{ocr-blur} & 0.997 & 0.006 & 47.66 & 0.343 & 37.9 & 0.5\% \\
\textsc{mm-pgd} & 0.989 & 0.035 & 45.51 & 0.035 & 37.8 & 61.3\% \\
\textsc{rand-noise} & 0.957 & 0.143 & 36.32 & 0.719 & 34.0 & 9.8\% \\
\textsc{geoshield} & 0.852 & 0.273 & 34.16 & 0.031 & 30.9 & 99.3\% \\
\textsc{gpt-image-2} & 0.861 & 0.140 & 27.44 & 0.826 & 47.0 & 99.8\% \\
\textbf{\name{}} & 0.920 & 0.109 & 28.67 & 0.786 & 40.9 & 19.7\% \\
\bottomrule
\end{tabular}
\end{table}

\section{Dataset Details}
\label{app:datasets}

\noindent\textbf{Cityscapes (geolocation).}
The 500-image set contains 25 images from each of 20 cities, drawn uniformly at random within each city from the \texttt{leftImg8bit} frames: Aachen, Bochum, Bremen, Cologne, Darmstadt, D\"usseldorf, Erfurt, Frankfurt, Hamburg, Hanover, Jena, Krefeld, Lindau, M\"onchengladbach, Strasbourg, Stuttgart, T\"ubingen, Ulm, Weimar, and Z\"urich (18 German cities plus Strasbourg and Z\"urich).
Stratifying by city keeps the set balanced, so that cities with many Cityscapes frames do not dominate the results.
Ground-truth coordinates come from the per-frame vehicle GPS metadata.
The 100-image subset used for perceptual quality, runtime, and the open-source oracle experiments contains 5 images per city from the same 20 cities.
All images are processed at their native $2048{\times}1024$ resolution, so fine text and small signs remain visible to both the defender and the attackers.

\noindent\textbf{Mapillary (geolocation holdout).}
The 200-image holdout set samples 20 cities worldwide with populations of at least 300{,}000 and 10 non-panoramic street-level images per city, drawn uniformly within a fixed radius of the city center.
It differs from Cityscapes in camera, viewpoint, and geography, and is used only for evaluation.

\noindent\textbf{CPRT (compositional privacy).}
We sample 500 images uniformly at random from the CPRT benchmark.
Table~\ref{tab:app-cprt} shows the distribution of their ground-truth risk scores.
The mean score is 0.484 (median 0.542), and 273 images (54.6\%) are \emph{high-concern} (ground-truth score $\ge 0.5$), so the sample covers both low- and high-risk images rather than only images with obvious identifiers.

\begin{table}[h]
\centering
\small
\caption{Ground-truth CPRT risk scores of the 500 sampled images.}
\label{tab:app-cprt}
\begin{tabular}{@{}lrrrrr@{}}
\toprule
\textbf{Score} & $[0,.2)$ & $[.2,.4)$ & $[.4,.6)$ & $[.6,.8)$ & $[.8,1]$ \\
\midrule
\textbf{Images} & 113 & 65 & 121 & 136 & 65 \\
\bottomrule
\end{tabular}
\end{table}

\noindent\textbf{PISC (social relationships).}
We sample 454 images with 1{,}109 annotated person pairs.
Table~\ref{tab:app-pisc} shows the label distribution under both granularities used in Sec.~\ref{sec:defense_model}.
Person boxes and IDs are drawn on the image before both attack and defense, so every condition refers to the same pairs.

\begin{table}[h]
\centering
\small
\caption{PISC pair labels in the evaluation set.}
\label{tab:app-pisc}
\begin{tabular}{@{}llr@{}}
\toprule
\textbf{3-way label} & \textbf{6-way label} & \textbf{Pairs} \\
\midrule
\multirow{3}{*}{Intimate (453)} & Friends & 307 \\
 & Family members & 109 \\
 & Couple & 37 \\
\midrule
\multirow{2}{*}{Non-intimate (426)} & Professional & 397 \\
 & Commercial & 29 \\
\midrule
No relationship (230) & No relationship & 230 \\
\midrule
\textbf{Total} & & 1{,}109 \\
\bottomrule
\end{tabular}
\end{table}

\section{Qualitative Examples}
\label{app:examples}

\newcommand{\geocell}[2]{\includegraphics[width=0.196\textwidth]{Latex_Jacky/figures/appendix/geo_#1_#2.jpg}}

\begin{figure*}[t]
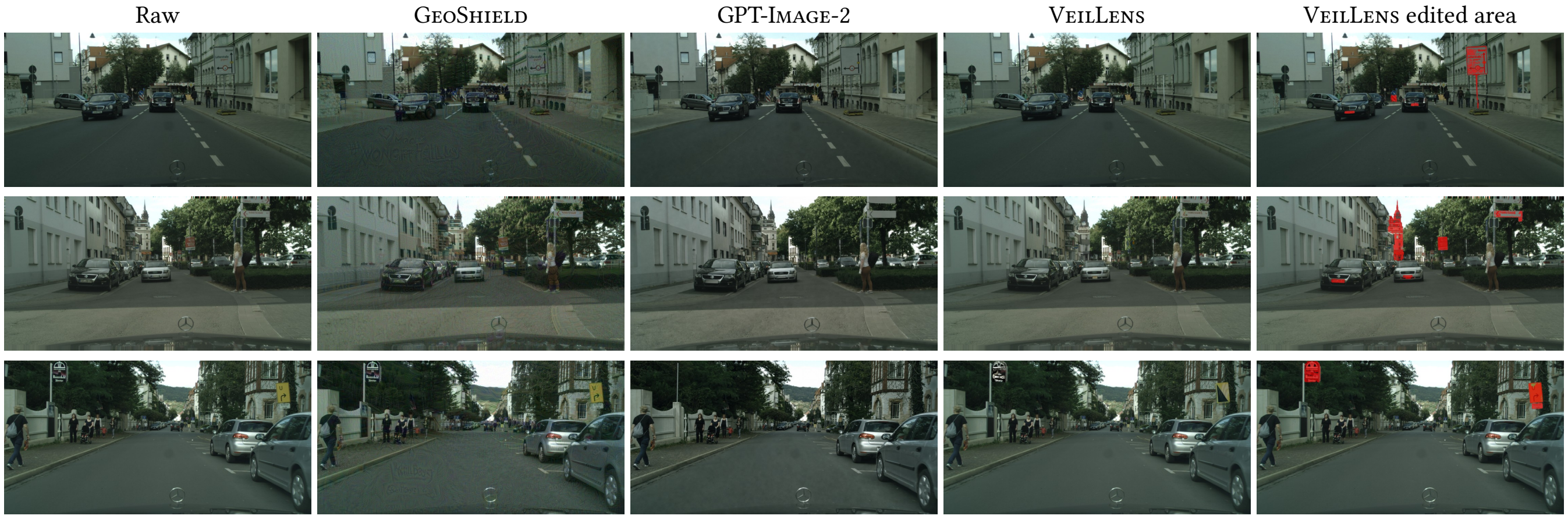

\centering
\setlength{\tabcolsep}{1pt}
\begin{tabular}{ccccc}
\small Raw & \small \textsc{GeoShield} & \small \textsc{GPT-Image-2} & \small \name{} & \small \name{} edited area \\
\geocell{jena23}{raw} & \geocell{jena23}{geoshield} & \geocell{jena23}{gptimage} & \geocell{jena23}{veillens} & \geocell{jena23}{diff} \\
\geocell{krefeld}{raw} & \geocell{krefeld}{geoshield} & \geocell{krefeld}{gptimage} & \geocell{krefeld}{veillens} & \geocell{krefeld}{diff} \\
\geocell{jena92}{raw} & \geocell{jena92}{geoshield} & \geocell{jena92}{gptimage} & \geocell{jena92}{veillens} & \geocell{jena92}{diff} \\
\end{tabular}
\caption{Geolocation defenses on three Cityscapes images. The last column marks in red the pixels that \name{} changed. Rows show a direction sign naming Weimar and the license plates (Jena, \name{} SSIM 0.992; 0.9\% of pixels changed); a church tower, an election poster and a direction sign naming Krefeld, and the license plates (Krefeld, SSIM 0.987; 1.6\%); and a pub sign advertising a regional beer and a detour sign (Jena, SSIM 0.991; 1.2\%). \textsc{GeoShield} perturbs the whole image and \textsc{GPT-Image-2} regenerates it, while \name{} leaves regions without grounded clues unchanged.}
\label{fig:app-geo-examples}
\Description{A grid of fifteen street images in three rows and five columns. Columns show the raw image, GeoShield, GPT-Image-2, VeilLens, and a red overlay of the pixels VeilLens changed. In every row the red overlay covers only a few signs, license plates, and, in the middle row, a church tower; the rest of each street scene is unchanged.}
\end{figure*}

\noindent\textbf{Geolocation.}
Figure~\ref{fig:app-geo-examples} compares the defenses on three Cityscapes images.
\name{} repaints only the grounded clues, such as signs naming a town or street, a pub sign advertising a regional beer, license plates, and a landmark church tower, and keeps the rest of the street scene unchanged.
Edits are not always this compact: when grounded clues cover large areas such as building facades or distinctive paving, the repainted area grows, and per-image perceptual cost is driven by masked area rather than by the number of rounds (App.~\ref{app:impl}).

\newcommand{\cprtcell}[2]{\includegraphics[width=0.48\columnwidth]{Latex_Jacky/figures/appendix/cprt_#1_#2.jpg}}

\begin{figure}[!htb]
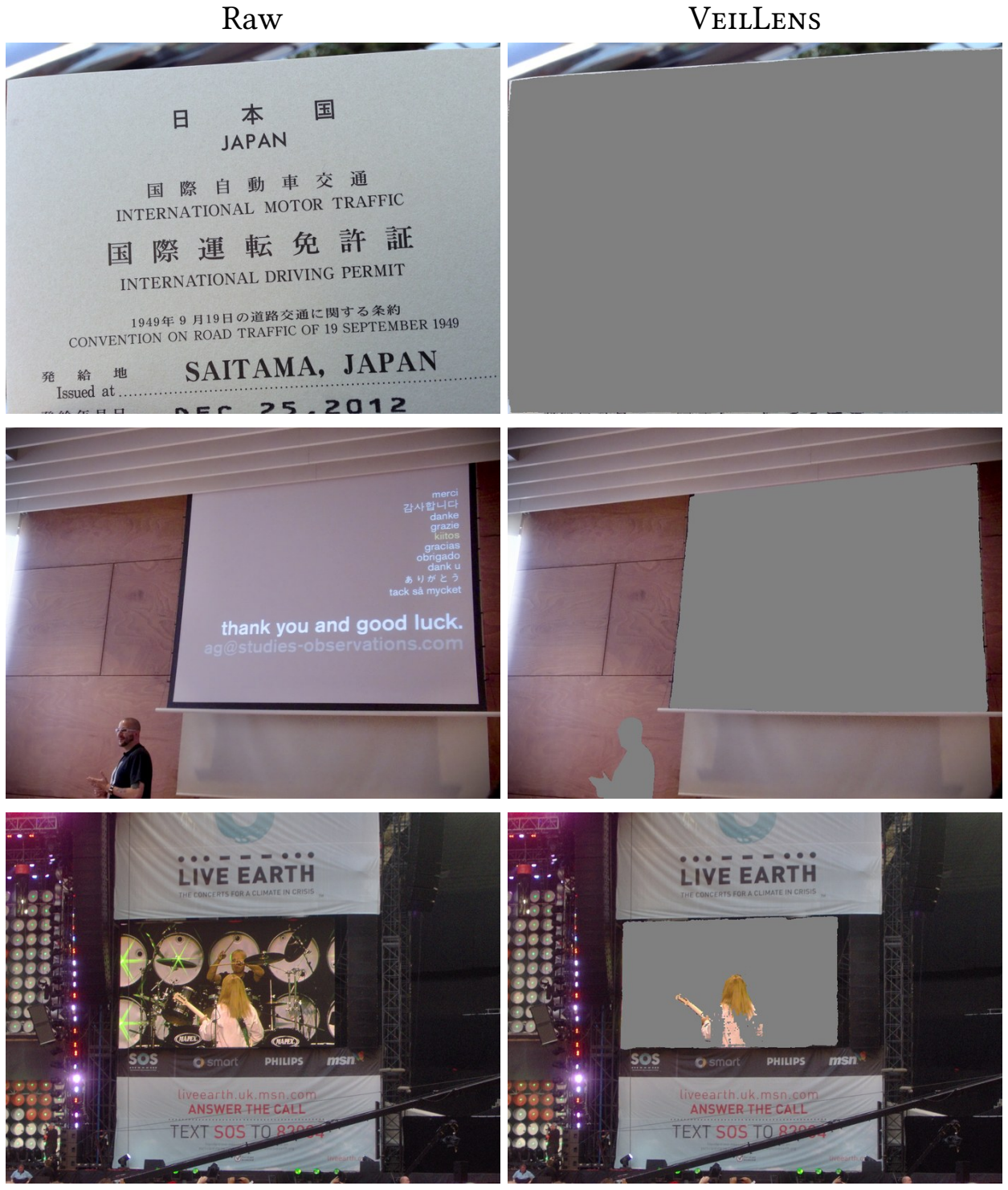

\centering
\setlength{\tabcolsep}{1pt}
\begin{tabular}{cc}
\small Raw & \small \name{} \\
\cprtcell{2017_16361409}{raw} & \cprtcell{2017_16361409}{veillens} \\
\cprtcell{2017_10201301}{raw} & \cprtcell{2017_10201301}{veillens} \\
\cprtcell{2017_14057640}{raw} & \cprtcell{2017_14057640}{veillens} \\
\end{tabular}
\caption{Compositional-privacy protection at CPRT Level~2 on three CPRT images (grey masks). Attacker risk scores (raw $\rightarrow$ protected, GPT-5.2 / Gemini-3.1-Flash-Lite): top, international driving permit, $0.74 \rightarrow 0.38 / 0.00$; middle, presentation slide with readable contact details, $0.56 \rightarrow 0.00 / 0.00$; bottom, concert video screen showing a performer's face, $0.44 \rightarrow 0.12 / 0.00$.}
\label{fig:app-cprt-examples}
\Description{Three pairs of images. Top: a scanned international driving permit, fully masked in grey after protection. Middle: a lecture hall with a projected slide showing text and an email address; the slide and the speaker are masked. Bottom: a concert stage whose central video screen, showing a drummer, is masked, while a guitarist seen from behind and the sponsor banners remain.}
\end{figure}·

\noindent\textbf{Compositional privacy.}
Figure~\ref{fig:app-cprt-examples} shows three CPRT images.
\name{} masks only the clues at or above the protection target: the identity document, the slide carrying readable contact details together with the speaker, and the video screen showing a performer's face.
Lower-level context such as the lecture hall and the event banners is kept, so the image still conveys the scene; the banners still reveal the event, which is a Level~3 location cue and therefore outside the Level~2 target.
These examples use grey masks to make the edited regions visible; with inpainting enabled, the same regions are repainted.

\clearpage %

\end{document}